\documentclass[submission,copyright]{eptcs}
\providecommand{\event}{VPT 2016}

\usepackage{proof}
\usepackage{amsmath}
\usepackage{fancybox}
\usepackage{stmaryrd}
\usepackage{amsthm}
\usepackage{amssymb}
\usepackage{enumerate}
\usepackage{enumitem}
\usepackage{pgf}
\usepackage{graphicx}

\usepackage{float}
\usepackage{keyval}
\usepackage{ifthen}
\usepackage{moreverb}

\newcommand{\doubleplus}{\ensuremath{+\!\!\!+\,}}
\newtheoremstyle{break}
  {\topsep}{\topsep}%
  {}{}%
  {\bfseries}{}%
  {\newline}{}%
\theoremstyle{break}
\newtheorem{definition}{Definition}[section]
\newtheorem{example}{Example}[section]
\newtheorem{property}{Property}[section]

\title{Program Transformation to Identify List-Based Parallel Skeletons}
\author{Venkatesh Kannan \hspace{15mm} G. W. Hamilton \vspace{2mm}
\institute{School of Computing, Dublin City University, Ireland}
\email{\{vkannan, hamilton\}@computing.dcu.ie}
}
\def\titlerunning{Program Transformation to Identify List-Based Parallel Skeletons}
\def\authorrunning{V. Kannan and G. W. Hamilton}

\begin{document}
\maketitle

\begin{abstract}
Algorithmic skeletons are used as building-blocks to ease the task of parallel programming by abstracting the details of parallel implementation from the developer. Most existing libraries provide implementations of skeletons that are defined over flat data types such as lists or arrays. However, skeleton-based parallel programming is still very challenging as it requires intricate analysis of the underlying algorithm and often uses inefficient intermediate data structures. Further, the algorithmic structure of a given program may not match those of list-based skeletons. In this paper, we present a method to automatically transform any given program to one that is defined over a list and is more likely to contain instances of list-based skeletons. This facilitates the parallel execution of a transformed program using existing implementations of list-based parallel skeletons. Further, by using an existing transformation called \textit{distillation} in conjunction with our method, we produce transformed programs that contain fewer inefficient intermediate data structures.
\end{abstract}

\section{Introduction}
\label{sec:introduction}
In today's computing systems, parallel hardware architectures that use multi-core CPUs and GPUs (Graphics Processor Units) are ubiquitous. On such hardware, it is essential that the programs developed be executed in parallel in order to effectively utilise the computing power that is available. To enable this, the parallelism that is inherent in a given program needs to be identified and exploited. However, parallel programming is tedious and error-prone when done by hand and is very difficult for a compiler to do automatically to the desired level.

To ease the task of parallel programming, a collection of algorithmic skeletons \cite{C1992AlgorithmicSkeletons, DFHKSW1993ParallelProgrammingUsing} are often used for program development to abstract away from the complexity of implementing the parallelism. In particular, \textit{map}, \textit{reduce} and \textit{zipWith} are primitive parallel skeletons that are often used for parallel programming \cite{MIEH2006ALibraryOfConstructiveSkeletons, GL2010SurveyOfAlgorithmicSkeletonFrameworks}. Most libraries such as Eden \cite{L2012Eden}, SkeTo \cite{MIEH2006SkeToLibrary}, Data Parallel Haskell (DPH) \cite{CLJKM2007DPHStatusReport}, and Accelerate \cite{CKLMG2011Accelerate} provide parallel implementations for these skeletons defined over flat data types such as lists or arrays. However, there are two main challenges in skeleton-based programming:
\begin{enumerate}
  \item Using multiple skeletons in a program often introduces inefficient intermediate data structures \cite{MKIHA2004AFusionEmbeddedSkeletonLibrary, MCKL2013OptimisingPurelyFunctionalGPUPrograms}.
  \item There may be a mismatch in data structures and algorithms used by the skeletons and the program \cite{ST1998ModelsAndLanguagesForParallelComputing, HTC1998ParallelisationInCalculationalForms}.
\end{enumerate}

For example, consider the matrix multiplication program shown in Example \ref{ex:matrix_multiplication_original_program}, where $mMul$ computes the product of two matrices $xss$ and $yss$. The function \textit{map} is used to compute the dot-product ($dotp$) of each row in $xss$ and those in the transpose of $yss$, which is computed by the function \textit{transpose}. Note that this definition uses multiple intermediate data structures, which is inefficient.
\begin{example}[Matrix Multiplication -- Original Program]
\label{ex:matrix_multiplication_original_program}
{\fontsize{10}{11}\selectfont
\begin{tabular}{@{\hspace{0mm}}l@{\hspace{2mm}}c@{\hspace{2mm}}l@{\hspace{0mm}}}
\multicolumn{3}{@{\hspace{0mm}}l@{\hspace{0mm}}}{$mMul\ ::\ [[a]]\ \to\ [[a]]\ \to\ [[a]]$} \\[2mm]
  \multicolumn{3}{@{\hspace{0mm}}l@{\hspace{0mm}}}{$mMul\ xss\ yss$} \\
  \multicolumn{3}{@{\hspace{0mm}}l@{\hspace{0mm}}}{$\mathbf{where}$} \\
  $mMul\ []\ yss$ & $=$ & $[]$ \\
  $mMul\ (xs:xss)\ yss$ & $=$ & $(map\ (transpose\ yss)\ (dotp\ xs)):(mMul\ xss\ yss)$ \\[1mm]
  $dotp\ xs\ ys$ & $=$ & $foldr\ (+)\ 0\ (zipWith\ (*)\ xs\ ys)$ \\[1mm]
  $transpose\ xss$ & $=$ & $transpose'\ xss\ []$ \\
  $transpose'\ []\ yss$ & $=$ & $yss$ \\
  $transpose'\ (xs:xss)\ yss$ & $=$ & $transpose'\ xss\ (rotate\ xs\ yss)$ \\[1mm]
  $rotate\ []\ yss$ & $=$ & $yss$ \\
  $rotate\ (x:xs)\ []$ & $=$ & $[x]:(rotate\ xs\ yss)$ \\
  $rotate\ (x:xs)\ (ys:yss)$ & $=$ & $(ys \doubleplus [x]):(rotate\ xs\ yss)$
\end{tabular}}
\end{example}
\noindent
A version of this program defined using the built-in \textit{map}, \textit{reduce} and \textit{zipWith} skeletons is shown in Example \ref{ex:hand_parallelised_matrix_multiplication}.
\begin{example}[Hand-Parallelised Matrix Multiplication]
\label{ex:hand_parallelised_matrix_multiplication}
{\fontsize{10}{11}\selectfont
\begin{tabular}{@{\hspace{0mm}}l@{\hspace{2mm}}c@{\hspace{2mm}}l@{\hspace{0mm}}}
  \multicolumn{3}{@{\hspace{0mm}}l@{\hspace{0mm}}}{$mMul\ xss\ yss$} \\
  \multicolumn{3}{@{\hspace{0mm}}l@{\hspace{0mm}}}{$\mathbf{where}$} \\
  $mMul\ []\ yss$ & $=$ & $[]$ \\
  $mMul\ (xs:xss)\ yss$ & $=$ & $(map\ (dotp\ xs)\ (transpose\ yss)):(mMul\ xss\ yss)$ \\[1mm]
  $dotp\ xs\ ys$ & $=$ & \textit{reduce}$\ (+)\ 0\ (zipWith\ (*)\ xs\ ys)$ \\[1mm]
  $transpose\ xss$ & $=$ & $transpose'\ xss\ []$ \\
  $transpose'\ []\ yss$ & $=$ & $yss$ \\
  $transpose'\ (xs:xss)\ yss$ & $=$ & $transpose'\ xss\ (rotate\ xs\ yss)$ \\[1mm]
  $rotate\ xs\ yss$ & $=$ & $zipWith\ (\lambda x. \lambda ys. (ys \doubleplus [x]))\ xs\ yss$
\end{tabular}}
\end{example}
As we can observe, though defined using parallel skeletons, this implementation still employs multiple intermediate data structures. For instance, the matrix constructed by the \textit{transpose} function is subsequently decomposed by \textit{map}. It is challenging to obtain a program that uses skeletons for parallel evaluation and contains very few intermediate data structures.

Therefore, it is desirable to have a method to automatically identify potential parallel computations in a given program, transform them to operate over flat data types to facilitate their execution using parallel skeletons provided in existing libraries, and reduce the number of inefficient intermediate data structures used.

In this paper, we present a transformation method with the following aspects:
\begin{enumerate}
  \item Reduces inefficient intermediate data structures in a given program using an existing transformation technique called \textit{distillation} \cite{HJ2012DistillationWithLTS}. (Section \ref{sec:distillation})
  \item Automatically transforms the distilled program by \textit{encoding} its inputs into a single \textit{cons}-list, referred to as the \textit{encoded list}. (Section \ref{sec:encoding_transformation})
  \item Allows for parallel execution of the encoded program using efficient implementations of map and map-reduce skeletons that operate over lists. (Section \ref{sec:parallel_execution_of_encoded_programs})
\end{enumerate}
In Section \ref{sec:evaluation}, we discuss the results of evaluating our proposed transformation method using two example programs. In Section \ref{sec:conclusion}, we present concluding remarks on possible improvements to our transformation method and discuss related work.

\section{Language}
\label{sec:language}
We focus on the automated parallelisation of functional programs because pure functional programs are free of side-effects, which makes them easier to analyse, reason about, and manipulate using program transformation techniques. This facilitates parallel evaluation of independent sub-expressions in a program. The higher-order functional language used in this work is shown in Definition \ref{def:language_grammar}.
\begin{definition}[Language Grammar]
\label{def:language_grammar}
{\fontsize{10}{11}\selectfont
  \begin{tabular}{@{\hspace{0mm}}r@{\hspace{2mm}}c@{\hspace{2mm}}l@{\hspace{0mm}}@{\hspace{42mm}}r@{\hspace{0mm}}}
    $\mathbf{data}\ T\ {\alpha}_1 \ldots {\alpha}_M$ & $::=$ & $c_1\ t^1_1 \ldots t^1_N\ | \ldots |\ c_K\ t^K_1 \ldots t^K_N$ & Type Declaration \\[1mm]
    $t$ & $::=$ & ${\alpha}_m\ |\ T\ t_1 \ldots t_M$ & Type Component
  \end{tabular}

  \vspace{3mm}
  \noindent
  \begin{tabular}{@{\hspace{0mm}}l@{\hspace{2mm}}c@{\hspace{2mm}}l@{\hspace{2mm}}r@{\hspace{0mm}}}
    $e$ & $::=$ & $x$ & Variable \\
        & $|$   & $c\ e_1 \ldots e_N$ & Constructor Application \\
        & $|$   & $e_0$ & Function Definition \\
        &       & $\mathbf{where}$ & \\
        &       & $f\ p^1_1 \ldots p^1_M\ x^1_{(M+1)} \ldots x^1_N = e_1\ \ldots\ f\ p^K_1 \ldots p^K_M\ x^K_{(M+1)} \ldots x^K_N = e_K$ & \\
        & $|$   & $f$ & Function Call \\
        & $|$   & $e_0\ e_1$ & Application \\
        & $|$   & $\mathbf{let}\ x_1 = e_1\ \ldots\ x_N = e_N\ \mathbf{in}\ e_0$ & $\mathbf{let}$--expression \\
        & $|$   & $\lambda x.e$ & $\lambda$--Abstraction \\
    $p$  & $::=$ & $x\ |\ c\ p_1 \ldots p_N$ & Pattern
  \end{tabular}}
\end{definition}

A program can contain data type declarations of the form shown in Definition \ref{def:language_grammar}. Here, $T$ is the name of the data type, which can be polymorphic, with type parameters ${\alpha}_1, \ldots, {\alpha}_M$. A data constructor $c_k$ may have zero or more components, each of which may be a type parameter or a type application. An expression $e$ of type $T$ is denoted by $e :: T$.

A program in this language can also contain an expression which can be a variable, constructor application, function definition, function call, application, \textbf{let}-expression or $\lambda$-expression. Variables introduced in a $\lambda$-expression, \textbf{let}-expression, or function definition are \textit{bound}, while all other variables are \textit{free}. Each constructor has a fixed arity. In an expression $c\ e_1 \ldots e_N$, $N$ must be equal to the arity of the constructor $c$. Patterns in a function definition header are grouped into two -- $p^k_1 \ldots p^k_M$ are inputs that are pattern-matched, and $x^k_{(M+1)} \ldots x^k_N$ are inputs that are not pattern-matched. The series of patterns $p^1_1 \ldots p^1_M,\ \ldots,\ p^K_1 \ldots p^K_M$ in a function definition must be non-overlapping and exhaustive. We use $[]$ and $(:)$ as short notations for the $Nil$ and $Cons$ constructors of a \textit{cons}-list and $\doubleplus$ for list concatenation. The set of free variables in an expression $e$ is denoted as $fv(e)$.

\begin{definition}[Context]
\label{def:context}
\normalfont
  A context $E$ is an expression with \textit{hole}s in place of sub-expressions. $E[e_1, \ldots, e_N]$ is the expression obtained by filling holes in context $E$ with the expressions $e_1, \ldots, e_N$.
\end{definition}

The call-by-name operational semantics of our language is defined using the one-step reduction relation shown in Definition \ref{def:one_step_reduction_relation}. 

\begin{definition}[One-Step Reduction Relation]
\label{def:one_step_reduction_relation}
\begin{tabular}{@{\hspace{0mm}}c@{\hspace{20mm}}c@{\hspace{20mm}}c@{\hspace{0mm}}}
  $\big( (\lambda x. e_0)\ e_1 \big) \overset{\beta}{\leadsto} \big( e_0 \{x \mapsto e_1\} \big)$ & $\infer{(e_0\ e_1) \overset{r}{\leadsto} (e'_0\ e_1)}{e_0 \overset{r}{\leadsto} e'_0}$ & $\infer{(e_0\ e_1) \overset{r}{\leadsto} (e_0\ e'_1)}{e_1 \overset{r}{\leadsto} e'_1}$ \\[5mm]
  \multicolumn{3}{@{\hspace{0mm}}c@{\hspace{0mm}}}{$\infer{(f\ e_1 \ldots e_N) \overset{f}{\leadsto} e \theta }{\big( f\ p_1 \ldots p_N = e \big) \wedge \big( \exists \theta \cdot \forall n \in \{1, \ldots, N\} \cdot e_n = p_n \theta \big)}$} \\[3mm]
  \multicolumn{3}{@{\hspace{0mm}}c@{\hspace{0mm}}}{$\big( \mathbf{let}\ x_1 = e_1\ \ldots\ x_N = e_N\ \mathbf{in}\ e_0 \big) \overset{\beta}{\leadsto} \big( e_0 \{x \mapsto e_1, \ldots, x_N \mapsto e_N\}\big)$}
\end{tabular}
\end{definition}

\section{Distillation}
\label{sec:distillation}
\noindent
\textbf{Objective:} A given program may contain a number of inefficient intermediate data structures. In order to reduce them, we use an existing transformation technique called \textit{distillation}.\\[2mm]
\textit{Distillation} \cite{HJ2012DistillationWithLTS} is a technique that transforms a program to remove intermediate data structures and yields a \textit{distilled program}. It is an unfold/fold-based transformation that makes use of well-known transformation steps -- unfold, generalise and fold \cite{PP1996RulesAndStrategies} -- and can potentially provide super-linear speedups to programs. The syntax of a distilled program $de^{\{\}}$ is shown in Definition \ref{def:distilled_form_grammar}. Here, $\rho$ is the set of variables introduced by $\mathbf{let}$--expressions; these are not decomposed by pattern-matching. Consequently, $de^{\{\}}$ is an expression that has fewer intermediate data structures.

\begin{definition}[Distilled Form Grammar]
\label{def:distilled_form_grammar}
{\fontsize{10}{11}\selectfont
  \begin{tabular}{@{\hspace{0mm}}r@{\hspace{2mm}}c@{\hspace{2mm}}l@{\hspace{50mm}}r@{\hspace{0mm}}}
    $de^{\rho}$ & $::=$ & $x\ de^{\rho}_1 \ldots de^{\rho}_N$ & Variable Application \\
                & $|$   & $c\ de^{\rho}_1 \ldots de^{\rho}_N$ & Constructor Application \\
                & $|$   & $de^{\rho}_0$ & Function Definition \\
                &       & $\mathbf{where}$ & \\
                &       & \multicolumn{2}{@{\hspace{0mm}}l@{\hspace{0mm}}}{$f\ p^1_1 \ldots p^1_M\ x^1_{(M+1)} \ldots x^1_N = de^{\rho}_1\ \ldots\ f\ p^K_1 \ldots p^K_M\ x^K_{(M+1)} \ldots x^K_N = de^{\rho}_K$} \\
                & $|$   & $f\ x_1 \ldots x_M\ x_{(M+1)} \ldots x_N$ & Function Application \\
                &       & s.t. $\forall x \in \{x_1, \ldots, x_M\} \cdot x \not\in \rho$ & \\
                &       & \multicolumn{2}{@{\hspace{10mm}}l@{\hspace{0mm}}}{$\forall n \in \{1, \ldots, N\} \cdot \big(x_n \in \rho\ \Rightarrow\ \forall k \in \{1, \ldots, K\} \cdot p^k_n = x^k_n\big)$} \\
                & $|$   & $\mathbf{let}\ x_1 = de^{\rho}_1\ \ldots\ x_N = de^{\rho}_N\ \mathbf{in}\ de^{\rho\ \cup\ \{x_1, \ldots, x_N\}}_1$ & $\mathbf{let}$--expression \\
                & $|$   & $\lambda x.de^{\rho}$ & $\lambda$--Abstraction \\
    $p$ & $::=$ & $x\ |\ c\ p_1 \ldots p_N$ & Pattern
  \end{tabular}}
\end{definition}

Example \ref{ex:matrix_multiplication_distilled_program} shows the distilled form of the example matrix multiplication program in Example \ref{ex:matrix_multiplication_original_program}. Here, we have lifted the definitions of functions $mMul_2$ and $mMul_3$ to the top level using lambda lifting for ease of presentation.

\begin{example}[Matrix Multiplication -- Distilled Program]
\label{ex:matrix_multiplication_distilled_program}
{\fontsize{10}{11}\selectfont
\begin{tabular}{@{\hspace{0mm}}l@{\hspace{2mm}}c@{\hspace{2mm}}l@{\hspace{0mm}}}
  \multicolumn{3}{@{\hspace{0mm}}l@{\hspace{0mm}}}{$mMul\ xss\ yss$} \\
  \multicolumn{3}{@{\hspace{0mm}}l@{\hspace{0mm}}}{$\mathbf{where}$} \\
  $mMul\ xss\ yss$ & $=$ & $mMul_1\ xss\ yss\ yss$ \\
  $mMul_1\ []\ zss\ yss$ & $=$ & $[]$ \\
  $mMul_1\ xss\ []\ yss$ & $=$ & $[]$ \\
  $mMul_1\ (xs:xss)\ (zs: zss)\ yss$ & $=$ & $\mathbf{let}\ v = \lambda xs. g\ xs$ \\
                                     &     & \begin{tabular}{@{\hspace{17mm}}l@{\hspace{2mm}}c@{\hspace{2mm}}l@{\hspace{0mm}}}
                                               \multicolumn{3}{@{\hspace{17mm}}l@{\hspace{0mm}}}{$\mathbf{where}$} \\
                                               $g\ []$ & $=$ & $0$ \\
                                               $g\ (x:xs)$ & $=$ & $x$
                                             \end{tabular} \\
                                     &     & $\mathbf{in}\ (mMul_2\ zs\ xs\ yss\ v):(mMul_1\ xss\ zss\ yss)$ \\
  $mMul_2\ []\ xs\ yss\ v$ & $=$ & $[]$ \\
  $mMul_2\ (z: zs)\ xs\ yss\ v$ & $=$ & $\mathbf{let}\ v' = \lambda xs. g\ xs$ \\
                                &     & \begin{tabular}{@{\hspace{18mm}}l@{\hspace{2mm}}c@{\hspace{2mm}}l@{\hspace{0mm}}}
                                          \multicolumn{3}{@{\hspace{18mm}}l@{\hspace{0mm}}}{$\mathbf{where}$} \\
                                          $g\ []$ & $=$ & $0$ \\
                                          $g\ (x:xs)$ & $=$ & $v\ xs$
                                        \end{tabular} \\
                                &     & $\mathbf{in}\ (mMul_3\ xs\ yss\ v):(mMul_2\ zs\ xs\ yss\ v')$ \\
  $mMul_3\ []\ yss\ v$ & $=$ & $0$ \\
  $mMul_3\ (x:xs)\ []\ v$ & $=$ & $0$ \\
  $mMul_3\ (x:xs)\ (ys:yss)\ v$ & $=$ & $(x * (v\ ys)) + (mMul_3\ xs\ yss\ v)$
\end{tabular}}
\end{example}

In this distilled program, function $mMul_1$ computes the product of matrices $xss$ and $yss$, and functions $mMul_2$ and $mMul_3$ compute the dot-product of a row in $xss$ and those in the transpose of $yss$. This version of matrix multiplication is free from intermediate data structures. In particular, distillation removes data structures that are constructed and subsequently decomposed as a part of the algorithm that is implemented in a given program.

\vspace{2mm}
\noindent
\textbf{Consequence:} Using the distillation transformation, we obtain a semantically equivalent version of the original program that has fewer intermediate data structures.

\section{Encoding Transformation}
\label{sec:encoding_transformation}
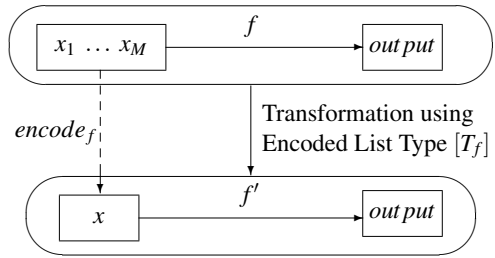
\begin{figure}[!ht]
\centering
\scalebox{0.85}{
  \begin{picture}(225,105)(0,0)
    \put(15,76){\framebox(57,20){$x_1\ \ldots\ x_M$}}
    \put(72,86){\vector(1,0){88}}
    \put(107,93){$f$}
    \put(160,76){\framebox(35,20){$output$}}
    \put(110,87){\oval(215,35)}

    \multiput(43,76)(0,-7){6}{\line(0,-1){4}}
    \put(43,34){\vector(0,-1){13}}
    \put(5,46){$encode_f$}

    \put(110,69){\vector(0,-1){40}}
    \put(115,53){Transformation using}
    \put(115,40){Encoded List Type $[T_f]$}

    \put(25,0){\framebox(35,20){$x$}}
    \put(60,10){\vector(1,0){100}}
    \put(105,17){$f'$}
    \put(160,2){\framebox(35,20){$output$}}
    \put(110,11){\oval(200,35)}
  \end{picture}}
\caption{Steps to Encode Inputs of Function $f$}
\label{fig:steps_to_encode_inputs_of_function_f}
\end{figure}

\noindent
\textbf{Objective:} The data types and the algorithm of a distilled program, which we want to parallelise, may not match with those of the skeletons defined over lists. This would inhibit the potential identification of parallel computations that could be encapsulated using the map or map-reduce skeletons. To resolve this, we define a transformation that encodes the inputs of a distilled program into a single \textit{cons}-list. The resulting encoded program is defined in a form that facilitates identification of list-based parallel skeleton instances.\\[2mm]
To perform the encoding transformation, we first lift the definitions of all functions in a distilled program to the top-level using lambda lifting. Following this, for each recursive function $f$ defined in the top-level \textbf{where}-expression of the distilled program, we encode the inputs $p_1, \ldots, p_M$ that are pattern-matched in the definition of $f$. Other inputs $x_{(M+1)}, \ldots, x_N$ that are never pattern-matched in the definition of $f$ are not encoded. Further, we perform this encoding only for the recursive functions in a distilled program because they are potential instances of parallel skeletons, which are also defined recursively. The three steps to encode inputs $x_1, \ldots, x_M$ of function $f$ into a \textit{cons}-list, referred to as the \textit{encoded list}, are illustrated in Figure \ref{fig:steps_to_encode_inputs_of_function_f} and described below. Here, we encode the pattern-matched inputs $x_1, \ldots, x_M$ into a \textit{cons}-list of type $[T_f]$, where $T_f$ is a new type created to contain the pattern-matched variables in $x_1, \ldots, x_M$.

Consider the definition of a recursive function $f$, with inputs $x_1, \ldots, x_M, x_{(M+1)}, \ldots, x_N$, of the form shown in Definition \ref{def:general_form_of_recursive_function_in_distilled_program_encoding_into_list} in a distilled program. Here, for each body $e_k$ corresponding to function header $f\ p^k_1 \ldots p^k_M\ x^k_{(M+1)} \ldots x^k_N$ in the definition of $f$, we use one of the recursive calls to function $f$ that may appear in $e_k$. All other recursive calls to $f$ in $e_k$ are a part of the context $E_k$.

\begin{definition}[General Form of Recursive Function in Distilled Program]
\label{def:general_form_of_recursive_function_in_distilled_program_encoding_into_list}
{\fontsize{10}{11}\selectfont
\begin{tabular}{@{\hspace{0mm}}l@{\hspace{2mm}}c@{\hspace{2mm}}l@{\hspace{0mm}}}
  \multicolumn{3}{@{\hspace{0mm}}l@{\hspace{0mm}}}{$f\ x_1 \ldots x_M\ x_{(M+1)} \ldots x_N$} \\
  \multicolumn{3}{@{\hspace{0mm}}l@{\hspace{0mm}}}{\textbf{where}} \\
  $f\ p^1_1 \ldots p^1_M\ x_{(M+1)} \ldots x_N$ & $=$ & $e_1$ \\ 
  $\vdots$ & & $\vdots$ \\
  $f\ p^K_1 \ldots p^K_M\ x_{(M+1)} \ldots x_N$ & $=$ & $e_K$ \\[1mm] 
  \multicolumn{3}{@{\hspace{0mm}}l@{\hspace{0mm}}}{where $\exists k \in \{1, \ldots, K\} \cdot e_k = E_k \left[ f\ x^k_1 \ldots x^k_M\ x^k_{(M+1)} \ldots x^k_{N} \right]$}
\end{tabular}}
\end{definition}
\noindent
The three steps to encode the pattern-matched inputs are as follows:
\begin{enumerate}[leftmargin=*]
  \item \textbf{Declare a new encoded data type $T_f$ :} \\
        First, we declare a new data type $T_f$ for elements of the encoded list. This new data type corresponds to the data types of the pattern-matched inputs of function $f$ that are encoded. The rules to declare type $T_f$ are shown in Definition \ref{def:rules_to_declare_encoded_input_data_type}.
        \begin{definition}[Rules to Declare Encoded Data Type for List]
        \label{def:rules_to_declare_encoded_input_data_type}
        {\fontsize{10}{11}\selectfont
        \begin{tabular}{@{\hspace{0mm}}l@{\hspace{2mm}}c@{\hspace{2mm}}l@{\hspace{0mm}}}
          $\mathbf{data}\ T_f\ {\alpha}_1 \ldots {\alpha}_G$ & $::=$ & $c_1\ T^1_1 \ldots T^1_L\ |\ \ldots\ |\ c_K\ T^K_1 \ldots T^K_L$ \\[0mm]
          \multicolumn{3}{@{\hspace{0mm}}l@{\hspace{0mm}}}{where} \\
          \multicolumn{3}{@{\hspace{0mm}}l@{\hspace{0mm}}}{${\alpha}_1, \ldots, {\alpha}_G$ are the type variables of the data types of the pattern-matched inputs} \\
          \multicolumn{3}{@{\hspace{0mm}}l@{\hspace{0mm}}}{$\forall k \in \{1, \ldots, K\} \cdot$} \\
          \multicolumn{3}{@{\hspace{0mm}}l@{\hspace{0mm}}}{$c_k$ is a fresh constructor for $T_f$ corresponding to $p^k_1 \ldots p^k_M$ of the pattern-matched inputs} \\
          \multicolumn{3}{@{\hspace{0mm}}l@{\hspace{0mm}}}{$\big\{(z_1 :: T^k_1), \ldots, (z_L :: T^k_L)\big\} = \left\{ \begin{tabular}{@{\hspace{0mm}}l@{\hspace{2mm}}l@{\hspace{0mm}}}
                                                                                                      $fv(E_k) \setminus \{x_{(M+1)}, \ldots, x_N\}$, & if $e_k = E_k \left[ f\ x^k_1 \ldots x^k_M\ x^k_{(M+1)} \ldots x^k_{N} \right]$ \\[1mm]
                                                                                                      $fv(e_k) \setminus \{x_{(M+1)}, \ldots, x_N\}$, & otherwise
                                                                                                    \end{tabular}\right.$} \\[5mm]
          \multicolumn{3}{@{\hspace{41mm}}l@{\hspace{0mm}}}{where $f\ p^k_1 \ldots p^k_M\ x_{(M+1)} \ldots x_N = e_k$}
        \end{tabular}}
        \end{definition}
        Here, a new constructor $c_k$ of the type $T_f$ is created for each set $p^k_1 \ldots p^k_M$ of the pattern-matched inputs $x_1 \ldots x_M$ of function $f$ that are encoded. As stated above, our objective is to encode the inputs of a recursive function $f$ into a list, where each element contains the pattern-matched variables consumed in an iteration of $f$. To achieve this, the variables bound by constructor $c_k$ correspond to the variables $z_1, \ldots, z_L$ in $p^k_1 \ldots p^k_M$ that occur in the context $E_k$ (if $e_k$ contains a recursive call to $f$) or the expression $e_k$ (otherwise). Consequently, the type components of constructor $c_k$ are the data types of the variables $z_1, \ldots, z_L$.
  \item \textbf{Define a function \textit{encode}$_f$ :} \\
        For a recursive function $f$ of the form shown in Definition \ref{def:general_form_of_recursive_function_in_distilled_program_encoding_into_list}, we use the rules in Definition \ref{def:rules_to_define_encode_f_function_for_list} to define function \textit{encode}$_f$ to build the encoded list, in which each element is of type $T_f$.
        \begin{definition}[Rules to Define Function \textit{encode}$_f$]
        \label{def:rules_to_define_encode_f_function_for_list}
        {\fontsize{10}{11}\selectfont
        \begin{tabular}{@{\hspace{0mm}}l@{\hspace{2mm}}c@{\hspace{2mm}}l@{\hspace{0mm}}}
          \multicolumn{3}{@{\hspace{0mm}}l@{\hspace{0mm}}}{$encode_f\ x_1 \ldots x_M$} \\
          \multicolumn{3}{@{\hspace{0mm}}l@{\hspace{0mm}}}{\textbf{where}} \\
          $encode_f\ p^1_1 \ldots p^1_{M}$ & $=$ & $e'_1$ \\ 
          $\vdots$ & & $\vdots$ \\
          $encode_f\ p^K_{1} \ldots p^K_{M}$ & $=$ & $e'_K$ \\[0mm] 
          \multicolumn{3}{@{\hspace{0mm}}l@{\hspace{0mm}}}{where} \\
          \multicolumn{3}{@{\hspace{0mm}}l@{\hspace{0mm}}}{$\forall k \in \{1, \ldots, K\} \cdot e'_k = \left\{ \begin{tabular}{@{\hspace{0mm}}l@{\hspace{2mm}}l@{\hspace{0mm}}}
            $\left[c_k\ z^k_1 \ldots z^k_L\right] \doubleplus (encode_f\ x^k_1 \ldots x^k_M)$, & if $e_k = E_k \left[ f\ x^k_1 \ldots x^k_M\ x^k_{(M+1)} \ldots x^k_{N} \right]$ \\[1mm]
            \multicolumn{2}{@{\hspace{0mm}}l@{\hspace{0mm}}}{where $\{z^k_1, \ldots, z^k_L\} = fv(E_k) \setminus \{x_{(M+1)}, \ldots, x_N\}$} \\[2mm]
            $\left[c_k\ z^k_1 \ldots z^k_L\right]$, & otherwise \\[1mm]
            \multicolumn{2}{@{\hspace{0mm}}l@{\hspace{0mm}}}{where $\{z^k_1, \ldots, z^k_L\} = fv(e_k) \setminus \{x_{(M+1)}, \ldots, x_N\}$}
          \end{tabular}\right.$} \\[9mm]
          \multicolumn{3}{@{\hspace{33mm}}l@{\hspace{0mm}}}{where $f\ p^k_1 \ldots p^k_M\ x_{(M+1)} \ldots x_N = e_k$}
        \end{tabular}}
        \end{definition}
        Here, for each pattern $p^k_1 \ldots p^k_M$ of the pattern-matched inputs, the \textit{encode}$_f$ function creates a list element. This element is composed of a fresh constructor $c_k$ of type $T_f$ that binds $z^k_1, \ldots, z^k_L$, which are the variables in $p^k_1 \ldots p^k_M$ that occur in the context $E_k$ (if $e_k$ contains a recursive call to $f$) or the expression $e_k$ (otherwise). The encoded input of the recursive call $f\ x^k_1 \ldots x^k_M\ x^k_{(M+1)} \ldots x^k_N$ is then computed by $encode_f\ x^k_1 \ldots x^k_M$ and appended to the element to build the complete encoded list for function $f$.
  \item \textbf{Transform the distilled program :} \\
        After creating the data type $T_f$ for the encoded list and the \textit{encode}$_f$ function for each recursive function $f$, we transform the distilled program using the rules in Definition \ref{def:rules_to_define_encoded_function_over_encoded_list} by defining a recursive function $f'$, which operates over the encoded list, corresponding to function $f$.
        
        \pagebreak
        \begin{definition}[Rules to Define Encoded Function Over Encoded List]
        \label{def:rules_to_define_encoded_function_over_encoded_list}
        {\fontsize{10}{11}\selectfont
        \begin{tabular}{@{\hspace{0mm}}l@{\hspace{2mm}}c@{\hspace{2mm}}l@{\hspace{0mm}}}
          \multicolumn{3}{@{\hspace{0mm}}l@{\hspace{0mm}}}{$f'\ x\ x_{(M+1)} \ldots x_N$} \\
          \multicolumn{3}{@{\hspace{0mm}}l@{\hspace{0mm}}}{\textbf{where}} \\
          $f'\ \big((c_1\ z^1_1 \ldots z^1_L):x^1\big)\ x_{(M+1)} \ldots x_N$ & $=$ & $e'_1$ \\
          $\vdots$ & & $\vdots$ \\
          $f'\ \big((c_K\ z^K_1 \ldots z^K_L):x^K\big)\ x_{(M+1)} \ldots x_N$ & $=$ & $e'_K$ \\
          \multicolumn{3}{@{\hspace{0mm}}l@{\hspace{0mm}}}{where} \\
          \multicolumn{3}{@{\hspace{0mm}}l@{\hspace{0mm}}}{$\forall k \in \{1, \ldots, K\} \cdot e'_k = \left\{ \begin{tabular}{@{\hspace{0mm}}l@{\hspace{2mm}}l@{\hspace{0mm}}}
                                                                                                                  $E_k \left[f'\ x^k\ x^k_{(M+1)} \ldots x^k_N\right]$, & if $e_k = E_k \left[ f\ x^k_1 \ldots x^k_M\ x^k_{(M+1)} \ldots x^k_{N} \right]$ \\
                                                                                                                  $e_k$, & otherwise
                                                                                                                \end{tabular}\right.$} \\[3mm]
          \multicolumn{3}{@{\hspace{33mm}}l@{\hspace{0mm}}}{where $f\ p^k_1 \ldots p^k_M\ x_{(M+1)} \ldots x_N = e_k$}
        \end{tabular}}
        \end{definition}
        \vspace{-2mm}
        Here, 
        \begin{itemize}
          \item In each function definition header of $f$, replace the pattern-matched inputs with a pattern to decompose the encoded list, such that the first element in the encoded list is matched with the corresponding pattern of the encoded type. For instance, a function header $f\ p_1 \ldots p_M\ x_{(M+1)} \ldots x_N$ is transformed to $f'\ p\ x_{(M+1)} \ldots x_N$, where $p$ is a pattern to match the first element in the encoded list with a pattern of the type $T_f$.
          \item In each call to function $f$, replace the pattern-matched inputs with their encoding. For instance, a call $f\ x_1 \ldots x_M\ x_{(M+1)} \ldots x_N$ is transformed to $f'\ x\ x_{(M+1)} \ldots x_N$, where $x$ is the encoding of the pattern-matched inputs $x_1, \ldots, x_M$.
        \end{itemize}
\end{enumerate}

The encoded data types, encode functions and encoded program obtained for the distilled matrix multiplication program from Example \ref{ex:matrix_multiplication_distilled_program} are shown in Example \ref{ex:matrix_multiplication_encoded_program}.

\begin{example}[Matrix Multiplication -- Encoded Program]
\label{ex:matrix_multiplication_encoded_program}
{\fontsize{10}{11}\selectfont
\begin{tabular}{@{\hspace{0mm}}l@{\hspace{2mm}}c@{\hspace{2mm}}l@{\hspace{0mm}}}
  $\textbf{data}\ T_{mMul_1}\ a$ & $::=$ & $c_1\ |\ c_2\ |\ c_3\ [a]\ [a]$ \\
  $\textbf{data}\ T_{mMul_2}\ a$ & $::=$ & $c_4\ |\ c_5$ \\
  $\textbf{data}\ T_{mMul_3}\ a$ & $::=$ & $c_6\ |\ c_7\ |\ c_8\ a\ [a]$
\end{tabular}

\vspace{0.5mm}
\noindent
\begin{tabular}{@{\hspace{0mm}}l@{\hspace{2mm}}c@{\hspace{2mm}}l@{\hspace{0mm}}}
  $encode_{mMul_1}\ []\ zss$ & $=$ & $[c_1]$ \\
  $encode_{mMul_1}\ xss\ []$ & $=$ & $[c_2]$ \\
  $encode_{mMul_1}\ (xs:xss)\ (zs:zss)$ & $=$ & $[c_3\ xs\ zs] \doubleplus (encode_{mMul_1}\ xss\ zss)$ \\[1mm]
  $encode_{mMul_2}\ []$ & $=$ & $[c_4]$ \\
  $encode_{mMul_2}\ (z:zs)$ & $=$ & $[c_5] \doubleplus (encode_{mMul_2}\ xs\ yss\ zs)$ \\[1mm]
  $encode_{mMul_3}\ []\ yss$ & $=$ & $[c_6]$ \\
  $encode_{mMul_3}\ (x:xs)\ []$ & $=$ & $[c_7]$ \\
  $encode_{mMul_3}\ (x:xs)\ (ys:yss)$ & $=$ & $[c_8\ x\ ys] \doubleplus (encode_{mMul_3}\ xs\ yss)$
\end{tabular}

\vspace{0.5mm}
\noindent
\begin{tabular}{@{\hspace{0mm}}l@{\hspace{2mm}}c@{\hspace{2mm}}l@{\hspace{0mm}}}
  \multicolumn{3}{@{\hspace{0mm}}l@{\hspace{0mm}}}{$mMul'\ xss\ yss$} \\
  \multicolumn{3}{@{\hspace{0mm}}l@{\hspace{0mm}}}{$\mathbf{where}$} \\
  $mMul'\ xss\ yss$ & $=$ & $mMul'_1\ (encode_{mMul_1}\ xss\ yss)\ yss$ \\
  $mMul'_1\ (c_1:w)\ yss$ & $=$ & $[]$ \\
  $mMul'_1\ (c_2:w)\ yss$ & $=$ & $[]$ \\
  $mMul_1\ \big((c_3\ xs\ zs):w\big)\ yss$ & $=$ & $\mathbf{let}\ v = \lambda xs. g\ xs$ \\
                                           &     & \begin{tabular}{@{\hspace{17mm}}l@{\hspace{2mm}}c@{\hspace{2mm}}l@{\hspace{0mm}}}
                                                     \multicolumn{3}{@{\hspace{17mm}}l@{\hspace{0mm}}}{$\mathbf{where}$} \\
                                                     $g\ []$ & $=$ & $0$ \\
                                                     $g\ (x:xs)$ & $=$ & $x$
                                                   \end{tabular} \\
                                           &     & $\mathbf{in}\ (mMul'_2\ (encode_{mMul_2}\ zs)\ xs\ yss\ v):(mMul'_1\ w\ yss)$
\end{tabular}

\noindent
\begin{tabular}{@{\hspace{0mm}}l@{\hspace{2mm}}c@{\hspace{2mm}}l@{\hspace{0mm}}}
  $mMul'_2\ (c_4:w)\ xs\ yss\ v$ & $=$ & $[]$ \\
  $mMul'_2\ (c_5:w)\ xs\ yss\ v$ & $=$ & $\mathbf{let}\ v' = \lambda xs. g\ xs$ \\
                                 &     & \begin{tabular}{@{\hspace{18mm}}l@{\hspace{2mm}}c@{\hspace{2mm}}l@{\hspace{0mm}}}
                                           \multicolumn{3}{@{\hspace{18mm}}l@{\hspace{0mm}}}{$\mathbf{where}$} \\
                                           $g\ []$ & $=$ & $0$ \\
                                           $g\ (x:xs)$ & $=$ & $v\ xs$
                                         \end{tabular} \\
                                 &     & $\mathbf{in}\ (mMul'_3\ (encode_{mMul_3}\ xs\ yss)\ v):(mMul'_2\ w\ xs\ yss\ v')$
\end{tabular}

\noindent
\begin{tabular}{@{\hspace{0mm}}l@{\hspace{2mm}}c@{\hspace{2mm}}l@{\hspace{0mm}}}
  $mMul'_3\ (c_6:w)\ v$ & $=$ & $0$ \\
  $mMul'_3\ (c_7:w)\ v$ & $=$ & $0$ \\
  $mMul'_3\ \big((c_8\ x\ ys):w\big)\ v$ & $=$ & $(x * (v\ ys)) + (mMul'_3\ w\ v)$
\end{tabular}}
\end{example}

\subsection{Correctness}
\label{sec:correctness}

The correctness of the encoding transformation can be established by proving that the result computed by each recursive function $f$ in the distilled program is the same as the result computed by the corresponding recursive function $f'$ in the encoded program. That is,
\begin{center}
  \begin{tabular}{l}
    $\big(f\ x_1 \ldots x_M\ x_{(M+1)} \ldots x_N\big) = \big(f'\ x\ x_{(M+1)} \ldots x_N\big)$ \\
    where $x = encode_f\ x_1 \ldots x_M$
  \end{tabular}
\end{center}
\noindent
\textbf{Proof:} \\
The proof is by structural induction over the encoded list type $[T_f]$. \\[2mm]
\textbf{Base Case:} \\
For the encoded list $x^k = \big((c_k\ z^k_1 \ldots z^k_L):[]\big)$ computed by $encode_f\ p^k_1 \ldots p^k_M$,
\begin{enumerate}[leftmargin=*]
  \item By Definition \ref{def:general_form_of_recursive_function_in_distilled_program_encoding_into_list}, L.H.S. evaluates to $e_k$.
  \item By Definition \ref{def:rules_to_define_encoded_function_over_encoded_list}, R.H.S. evaluates to $e_k$.
\end{enumerate}
\textbf{Inductive Case:} \\
For the encoded list $x^k = \big((c_k\ z^k_1 \ldots z^k_L):x^k\big)$ computed by $encode_f\ p^k_1 \ldots p^k_M$,
\begin{enumerate}[leftmargin=*]
  \item By Definition \ref{def:general_form_of_recursive_function_in_distilled_program_encoding_into_list}, L.H.S. evaluates to $E_k \left[f\ x^k_1 \ldots x^k_M\ x^k_{(M+1)} \ldots x^k_N\right]$.
  \item By Definition \ref{def:rules_to_define_encoded_function_over_encoded_list}, R.H.S. evaluates to $E_k \left[f'\ x^k\ x^k_{(M+1)} \ldots x^k_N\right]$.
  \item By inductive hypothesis, $\big(f\ x_1 \ldots x_M\ x_{(M+1)} \ldots x_N\big) = \big(f'\ x\ x_{(M+1)} \ldots x_N\big)$. \qed
\end{enumerate}

\vspace{3mm}
\noindent
\textbf{Consequence:} As a result of the encoding transformation, the pattern-matched inputs of a recursive function are encoded into a \textit{cons}-list by following the recursive structure of the function. Parallelisation of the encoded program produced by this transformation by identifying potential instances of \textit{map} and \textit{map-reduce} skeletons is discussed in Section \ref{sec:parallel_execution_of_encoded_programs}.

\section{Parallel Execution of Encoded Programs}
\label{sec:parallel_execution_of_encoded_programs}
\noindent
\textbf{Objective:} An encoded program defined over an encoded list is more likely to contain recursive functions that resemble the structure of \textit{map} or \textit{map-reduce} skeletons. This is because the $encode_f$ function constructs the encoded list in such a way that it reflects the recursive structure of the \textit{map} and \textit{map-reduce} skeletons defined over a \textit{cons}-list. Therefore, we look for instances of these skeletons in our encoded program.\\[2mm]
In this work, we identify instances of only \textit{map} and \textit{map-reduce} skeletons in an encoded program. This is because, as shown in Property \ref{pro:non_associative_reduction_operator_for_encoded_list}, any function that is an instance of a reduce skeleton in an encoded program that operates over an encoded list cannot be efficiently evaluated in parallel because the reduction operator will not be associative.

\pagebreak
\begin{property}[Non-Associative Reduction Operator for Encoded List]
\label{pro:non_associative_reduction_operator_for_encoded_list}
Given an encoded program defined over an encoded list, the reduction operator $\oplus$ in any instance of a reduce skeleton is not associative, that is $\forall x, y, z \cdot (x \oplus (y \oplus z)) \neq ((x \oplus y) \oplus z)$. \\[7mm]
\textbf{Proof:}
\begin{enumerate}
  \item From Definition \ref{def:rules_to_define_encoded_function_over_encoded_list}, given an encoded function $f'$,
        \begin{center}
        \begin{tabular}{@{\hspace{0mm}}l@{\hspace{0mm}}}
          $f'\ ::\ [T_f]\ \to\ T_{(M+1)}\ \ldots\ T_N\ \to\ b$ \\[1mm]
          where $[T_f]$ is the encoded list data type. \\
          \hspace{10mm}$T_{(M+1)}, \ldots, T_N$ are data types for inputs that are not encoded. \\
          \hspace{10mm}$b$ is the output data type.
        \end{tabular}
        \end{center}
  \item If $f'$ is an instance of a reduce skeleton, then the type of the binary reduction operator is given by $\oplus :: T_f \to b \to b$.
  \item Given that $T_f$ is a newly created data type, it follows from (2) that the binary operator $\oplus$ is not associative because the two input data types $T_f$ and $b$ cannot not be equal. \qed
\end{enumerate}
\end{property}

\subsection{Identification of Skeletons}
\label{sec:identification_of_skeletons}
To identify skeleton instances in a given program, we use a framework of \textit{labelled transition systems (LTSs)}, presented in Definition \ref{def:labelled_transition_system}, to represent and analyse the encoded programs and skeletons. This is because LTS representations enable matching the recursive structure of the encoded program with that of the skeletons rather than finding instances by matching expressions.

\begin{definition}[Labelled Transition System (LTS)]
\label{def:labelled_transition_system}
{\fontsize{10}{11}\selectfont
A LTS for a given program is given by $l = (\mathcal{S},\ s_0,\ Act,\ \to)$ where:
  \begin{itemize}
  \setlength\itemsep{0mm}
    \item $\mathcal{S}$ is the set of \textit{states} of the LTS, where each state has a unique label $s$.
    \item $s_0 \in \mathcal{S}$ is the start state denoted by \textit{start(l)}.
    \item \textit{Act} is one of the following actions:
    \begin{itemize}
    \setlength\itemsep{0mm}
      \item $x$, a free variable or \textbf{let}-expression variable,
      \item $c$, a constructor in an application,
      \item $\lambda$, a $\lambda$-abstraction,
      \item $@$, an expression application,
      \item $\#i$, the $i^{th}$ argument in an application,
      \item $p$, the set of patterns in a function definition header,
      \item \textbf{let}, a \textbf{let}-expression body.
    \end{itemize}
    \item $\to\ \subseteq\ \mathcal{S} \times Act \times \mathcal{S}$ relates pairs of states by actions in \textit{Act} such that if $s \in \mathcal{S}$ and $s \xrightarrow{\alpha} s'$ then $s' \in \mathcal{S}$ where $\alpha \in Act$.
  \end{itemize}}
\end{definition}

The LTS corresponding to a given program $e$ can be constructed by $\mathcal{L} \llbracket e \rrbracket\ s_0\ \emptyset\ \emptyset$ using the rules $\mathcal{L}$ shown in Definition \ref{def:lts_representation_of_program}. Here, $s_0$ is the start state, $\phi$ is the set of previously encountered function calls mapped to their corresponding states, and $\Delta$ is the set of function definitions. A LTS built using these rules is always finite because if a function call is re-encountered, then the corresponding state is reused.
\begin{definition}[LTS Representation of Program]
\label{def:lts_representation_of_program}
{\fontsize{10}{11}\selectfont
\begin{tabular}{@{\hspace{0mm}}l@{\hspace{1mm}}c@{\hspace{1mm}}l@{\hspace{0mm}}}
  $\mathcal{L} \llbracket x \rrbracket\ s\ \phi\ \Delta$ & $=$ & $s \to (x,\mathbf{0})$ \\
  $\mathcal{L} \llbracket c\ e_1 \ldots e_N \rrbracket\ s\ \phi\ \Delta$ & $=$ & $s \to (c, \mathbf{0}), (\#1, \mathcal{L} \llbracket e_1 \rrbracket\ s_1\ \phi\ \Delta), \ldots, (\#N, \mathcal{L} \llbracket e_N \rrbracket\ s_N\ \phi\ \Delta)$ \\
  $\mathcal{L} \llbracket e_0\ \mathbf{where}\ \delta_1 \ldots \delta_J \rrbracket\ s\ \phi\ \Delta$ & $=$ & $\mathcal{L} \llbracket e_0 \rrbracket\ s\ \phi\ \big( \Delta \cup \{f_1 \mapsto \delta_1, \ldots, f_J \mapsto \delta_J\} \big)$ \\
  \multicolumn{3}{@{\hspace{0mm}}r@{\hspace{0mm}}}{where $\forall j \in \{1, \ldots, J\} \cdot \delta_j = f_j\ p^1_1 \ldots p^1_M\ x^1_{(M+1)} \ldots x^1_N = e_1\ \ldots\ f_j\ p^K_1 \ldots p^K_M\ x^K_{(M+1)} \ldots x^K_N = e_K$} \\
  $\mathcal{L} \llbracket f \rrbracket\ s\ \phi\ \Delta$ & $=$ & $\left \{ \begin{tabular}{@{\hspace{1mm}}l@{\hspace{3mm}}l@{\hspace{0mm}}}
                                                                             $l$ \hfill where $\phi(f) = start(l)$, & if $f \in dom(\phi)$ \\
                                                                             $\mathcal{L} \llbracket \Delta(f) \rrbracket\ s\ (\phi \cup \{f \mapsto s\})\ \Delta$, & otherwise
                                                                           \end{tabular} \right .$ \\
  $\mathcal{L} \left\llbracket \begin{tabular}{@{\hspace{0mm}}l@{\hspace{1mm}}c@{\hspace{1mm}}l@{\hspace{0mm}}}
                                 $f\ p^1_1 \ldots p^1_M\ x^1_{(M+1)} \ldots x^1_N$ & $=$ & $e_1$ \\
                                 \hspace{2mm}$\vdots$ & & $\vdots$ \\
                                 $f\ p^K_1 \ldots p^K_M\ x^K_{(M+1)} \ldots x^K_N$ & $=$ & $e_K$
                               \end{tabular}\right\rrbracket s\ \phi\ \Delta$
  & $=$ & $\left\{\begin{tabular}{@{\hspace{0mm}}l@{\hspace{0mm}}}
                    $s \to (p^1_1 \ldots p^1_M\ x^1_{(M+1)} \ldots x^1_N, l_1),\ \ldots,$ \\
                    \hspace{7mm}$(p^K_1 \ldots p^K_M\ x^K_{(M+1)} \ldots x^K_N, l_K)$ \\[1mm]
                    where $\forall k \in \{1, \ldots, K\} \cdot l_k = \big(\mathcal{L} \llbracket e_k \rrbracket\ s_k\ \phi\ \Delta \big)$
                  \end{tabular}\right.$ \\
  $\mathcal{L} \llbracket e_0\ e_1 \rrbracket\ s\ \phi\ \Delta$ & $=$ & $s \to (@, \mathcal{L} \llbracket e_0 \rrbracket\ s_0\ \phi\ \Delta), (\#1, \mathcal{L} \llbracket e_1 \rrbracket\ s_1\ \phi\ \Delta)$ \\
  $\mathcal{L} \llbracket \mathbf{let}\ x_1 = e_1\ \ldots\ x_N\ = e_N\ \mathbf{in}\ e_0 \rrbracket\ s\ \phi\ \Delta$ & $=$ & $s \to \big(\mathbf{let}, \mathcal{L} \llbracket e_0 \rrbracket\ s_0\ \phi\ \Delta),$ \\
  & & \hspace{7mm}$(x_1, \mathcal{L} \llbracket e_1 \rrbracket\ s_1\ \phi\ \Delta), \ldots, (x_N, \mathcal{L} \llbracket e_N \rrbracket\ s_N\ \phi\ \Delta)$ \\
  $\mathcal{L} \llbracket \lambda x. e \rrbracket\ s\ \phi\ \Delta$ & $=$ & $s \to (\lambda, \mathcal{L} \llbracket e \rrbracket\ s_1\ \phi\ \Delta)$
\end{tabular}}
\end{definition}

\begin{definition}[LTS Substitution]
\label{def:lts_substitution}
\normalfont
A substitution is denoted by $\theta = \{x_1 \mapsto l_1, \ldots, x_N \mapsto l_N\}$. If $l$ is an LTS, then $l\theta = l\{x_1 \mapsto l_1, \ldots, x_N \mapsto l_N\}$ is the result of simultaneously replacing the LTSs $s_n \rightarrow (x_n,\mathbf{0})$ with the corresponding LTS $l_n$ in the LTS $l$ while ensuring that bound variables are renamed appropriately to avoid name capture.
\end{definition}

Potential instances of skeleton LTSs can be identified and replaced with suitable calls to corresponding skeletons in the LTS of an encoded program $l$ by $\mathcal{S} \llbracket l \rrbracket\ \emptyset\ \langle \rangle\ \omega$ using the rules presented in Definition \ref{def:extraction_of_program_from_lts_with_skeletons}.

\begin{definition}[Extraction of Program from LTS with Skeletons]
\label{def:extraction_of_program_from_lts_with_skeletons}
{\fontsize{10}{11}\selectfont
\begin{tabular}{@{\hspace{0mm}}l@{\hspace{2mm}}c@{\hspace{2mm}}l@{\hspace{0mm}}}
  $\mathcal{S} \llbracket l \rrbracket\ \rho\ \sigma\ \omega$ & $=$ & $\left \{ \begin{tabular}{@{\hspace{0mm}}l@{\hspace{2mm}}l@{\hspace{0mm}}}
                                                                                  $\left.\begin{tabular}{@{\hspace{0mm}}l@{\hspace{2mm}}l@{\hspace{1mm}}}
                                                                                           $f\ e_1 \ldots e_N$, & if $\exists (f\ x_1 \ldots x_N, l') \in \omega, \theta \cdot l' \theta = l$ \\
                                                                                           \multicolumn{2}{@{\hspace{0mm}}l@{\hspace{0mm}}}{where} \\
                                                                                           \multicolumn{2}{@{\hspace{0mm}}l@{\hspace{0mm}}}{$\theta = \{x_1 \mapsto l_1, \ldots, x_N \mapsto l_N\}$} \\
                                                                                           \multicolumn{2}{@{\hspace{0mm}}l@{\hspace{0mm}}}{$\forall n \in \{1, \ldots, N\} \cdot e_n = (\mathcal{S} \llbracket l_n \rrbracket\ \rho\ \sigma\ \omega)$} \\[2mm]
                                                                                           $f\ e_1 \ldots e_N$, & if $\exists (s,f) \in \rho \cdot start(l) = s$ \\
                                                                                           \multicolumn{2}{@{\hspace{0mm}}l@{\hspace{0mm}}}{where $\sigma = \langle e_1, \ldots, e_N \rangle$} \\[2mm]
                                                                                           $\mathcal{S'} \llbracket l \rrbracket\ \rho\ \sigma\ \omega$, & otherwise
                                                                                         \end{tabular}\right\}$, & if $\exists s \in states(l), \alpha \cdot s \overset{\alpha}{\to} start(l)$ \\[15mm]
                                                                                  $\mathcal{S'} \llbracket l \rrbracket\ \rho\ \sigma\ \omega$, & otherwise
                                                                                \end{tabular} \right .$
\end{tabular}

\noindent
\begin{tabular}{@{\hspace{0mm}}l@{\hspace{1mm}}c@{\hspace{1mm}}l@{\hspace{0mm}}}
  $\mathcal{S'} \llbracket s \to (x, \mathbf{0}) \rrbracket\ \rho\ \sigma\ \omega$ & $=$ & $x\ e_1 \ldots e_N$ \hspace{5mm} where $\sigma = \langle e_1, \ldots, e_N \rangle$ \\
  $\mathcal{S'} \llbracket s \to (c, \mathbf{0}), (\#1, l_1), \ldots, (\#N, l_N) \rrbracket\ \rho\ \sigma\ \omega$ & $=$ & $c\ (\mathcal{S} \llbracket l_1 \rrbracket\ \rho\ \sigma\ \omega) \ldots (\mathcal{S} \llbracket l_N \rrbracket\ \rho\ \sigma\ \omega)$ \\
  $\mathcal{S'} \left\llbracket \begin{tabular}{@{\hspace{0mm}}l@{\hspace{0mm}}}
                                  $s \to (p^1_1 \ldots p^1_M\ x^1_{(M+1)} \ldots x^1_N, l_1),\ \ldots,$ \\
                                  \hspace{7mm}$(p^K_1 \ldots p^K_M\ x^K_{(M+1)} \ldots x^K_N, l_K)$
                                \end{tabular}\right\rrbracket \rho\ \sigma\ \omega$
  & $=$ & $\left\{ \begin{tabular}{@{\hspace{0mm}}l@{\hspace{2mm}}c@{\hspace{2mm}}l@{\hspace{0mm}}}
                     \multicolumn{3}{@{\hspace{0mm}}l@{\hspace{0mm}}}{$f\ e_1 \ldots e_N$} \\
                     \multicolumn{3}{@{\hspace{0mm}}l@{\hspace{0mm}}}{\textbf{where}} \\
                     $f\ p^1_1 \ldots p^1_M\ x^1_{(M+1)} \ldots x^1_N$ & $=$ & $e'_1$ \\
                     \hspace{2mm}$\vdots$ & & $\vdots$ \\
                     $f\ p^K_1 \ldots p^K_M\ x^K_{(M+1)} \ldots x^K_N$ & $=$ & $e'_K$ \\[1mm]
                     \multicolumn{3}{@{\hspace{0mm}}l@{\hspace{0mm}}}{where $f$ is fresh, $\sigma = \langle e_1, \ldots, e_N \rangle$} \\
                     \multicolumn{3}{@{\hspace{9mm}}l@{\hspace{0mm}}}{$\forall k \in \{1, \ldots, K\} \cdot e'_k = \big( \mathcal{S} \llbracket l_k \rrbracket\ \rho'\ \langle \rangle\ \omega \big)$} \\
                     \multicolumn{3}{@{\hspace{9mm}}l@{\hspace{0mm}}}{$\rho' = \rho \cup \{(s, f)\}$}
                   \end{tabular}\right.$ \\[16mm]
  $\mathcal{S'} \llbracket s \to (@, l_0), (\#1, l_1) \rrbracket\ \rho\ \sigma\ \omega$ & $=$ & $\mathcal{S} \llbracket l_0 \rrbracket\ \rho\ \langle(\mathcal{S} \llbracket l_1 \rrbracket\ \rho\ \omega\ \langle \rangle):\sigma\rangle\ \omega$ \\
  $\mathcal{S'} \llbracket s \to (\mathbf{let}, l_0), (x_1, l_1), \ldots, (x_N, l_N) \rrbracket\ \rho\ \sigma\ \omega$ & $=$ & $\mathbf{let}\ x_1 = (\mathcal{S} \llbracket l_1 \rrbracket\ \rho\ \sigma\ \omega)\ \ldots\ x_N = (\mathcal{S} \llbracket l_N \rrbracket\ \rho\ \sigma\ \omega)$ \\
  & & $\mathbf{in}\ (\mathcal{S} \llbracket l_0 \rrbracket\ \rho\ \sigma\ \omega)$ \\
  $\mathcal{S'} \llbracket s \to (\lambda, l) \rrbracket\ \rho\ \sigma\ \omega$ & $=$ & $\lambda x. (\mathcal{S} \llbracket l \rrbracket\ \rho\ \sigma\ \omega)$ \hspace{5mm} where $x$ is fresh
\end{tabular}}
\end{definition}

Here, the parameter $\rho$ contains the set of new functions that are created and associates them with their corresponding states in the LTS. The parameter $\sigma$ contains the sequence of arguments of an application expression. The set $\omega$ is initialised with pairs of application expression and corresponding LTS representation of each parallel skeleton to be identified in a given LTS; for example, $(map\ xs\ f, l)$ is a pair in $\omega$ where $map\ xs\ f$ is the application expression for $map$ and $l$ is its LTS representation.

The definitions of list-based map and map-reduce skeletons whose instances we identify in an encoded program are as follows:
\begin{flushleft}
\begin{tabular}{@{\hspace{0mm}}l@{\hspace{1mm}}c@{\hspace{1mm}}l@{\hspace{0mm}}}
  $map$ & $::$ & $[a] \to (a \to b) \to [b]$ \\
  $map\ []\ f$ & $=$ & $[]$ \\
  $map\ (x:xs)\ f$ & $=$ & $(f\ x):(map\ xs\ f)$ 
\end{tabular}

\noindent
\begin{tabular}{@{\hspace{0mm}}l@{\hspace{1mm}}c@{\hspace{1mm}}l@{\hspace{0mm}}}
  $mapReduce$ & $::$ & $[a] \to (b \to b \to b) \to b \to (a \to b) \to b$ \\
  $mapReduce\ []\ g\ v\ f$ & $=$ & $v$ \\
  $mapReduce\ (x:xs)\ g\ v\ f$ & $=$ & $g\ (f\ x)\ (mapReduce\ xs\ g\ v\ f)$
\end{tabular}
\end{flushleft}

\begin{property}[Non-Empty Encoded List]
\label{pro:non_empty_encoded_list}
Given rules in Definition \ref{def:rules_to_define_encode_f_function_for_list} to encode inputs into a list, $\forall f, x_1, \ldots, x_M \cdot \big(encode_f\ x_1 \ldots x_M\big) \neq []$. \\
\textbf{Proof:} \\
From Definition \ref{def:rules_to_define_encode_f_function_for_list}, $\exists k \in \{1, \ldots, K\} \cdot p^k_1 \ldots p^k_M$ that matches inputs $x_1 \ldots x_M$.\\ Consequently, $\big(encode_f\ x_1 \ldots x_M\big) = [c_k\ z^k_1 \ldots z^k_L] \doubleplus \big(encode_f\ x^k_1 \ldots x^k_M\big)$. Therefore, the list computed by $encode_f\ x_1 \ldots x_M$ is at least a singleton.\qed
\end{property}

From Property \ref{pro:non_empty_encoded_list}, it is evident that the encoded programs produced by our transformation will always be defined over non-empty encoded list inputs. Consequently, to identify instances of \textit{map} and \textit{mapReduce} skeletons in an encoded program, we represent only the patterns corresponding to non-empty inputs, i.e. $(x:xs)$, in the LTSs built for the skeletons.

As an example, the LTSs built for the map skeleton and the $mMul'_1$ function in the encoded program for matrix multiplication in Example \ref{ex:matrix_multiplication_encoded_program} are illustrated in Figures \ref{fig:lts_for_map_skeleton} and \ref{fig:lts_for_mmul'_1_function}, respectively. Here, we observe that the LTS of $mMul'_1$ is an instance of the LTS of \textit{map} skeleton. Similarly, the LTS of $mMul'_3$ is an instance of the LTS of \textit{mapReduce} skeleton.

\begin{figure}[!ht]
\centering
\begin{minipage}[b]{0.4\textwidth}
\scalebox{0.99}{
\begin{pgfpicture}{0cm}{0cm}{5cm}{8cm}
\pgfsetendarrow{\pgfarrowto}
\pgfnodecircle{node1}[stroke]{\pgfxy(3.5,6.5)}{3pt}
\pgfnodecircle{node11}[stroke]{\pgfxy(3.0,5.75)}{3pt}
  \pgfnodeconnline{node1}{node11}
  \pgfnodelabel{node1}{node11}[0.5][0pt]{\pgfbox[center,center]{\small{$@$}}}
  \pgfnodecircle{node111}[stroke]{\pgfxy(2.5,5.0)}{3pt}
    \pgfnodeconnline{node11}{node111}
    \pgfnodelabel{node11}{node111}[0.5][0pt]{\pgfbox[center,center]{\small{$@$}}}
    \pgfnodecircle{node1111}[stroke]{\pgfxy(2.0,4.25)}{3pt}
      \pgfnodeconnline{node111}{node1111}
      \pgfnodelabel{node111}{node1111}[0.5][0pt]{\pgfbox[center,center]{\small{$(x:xs)\ f$}}}
      \pgfnodecircle{node11111}[stroke]{\pgfxy(1.0,3.5)}{3pt}
        \pgfnodeconnline{node1111}{node11111}
        \pgfnodelabel{node1111}{node11111}[0.5][0pt]{\pgfbox[right,center]{\small{$(:)$}}}
      \pgfnodecircle{node11112}[stroke]{\pgfxy(2.0,3.5)}{3pt}
        \pgfnodeconnline{node1111}{node11112}
        \pgfnodelabel{node1111}{node11112}[0.5][0pt]{\pgfbox[center,center]{\small{$\#1$}}}
        \pgfnodecircle{node111121}[stroke]{\pgfxy(1.5,2.75)}{3pt}
          \pgfnodeconnline{node11112}{node111121}
          \pgfnodelabel{node11112}{node111121}[0.5][0pt]{\pgfbox[center,center]{\small{$@$}}}
          \pgfnodecircle{node1111211}[stroke]{\pgfxy(1.5,2.0)}{3pt}
            \pgfnodeconnline{node111121}{node1111211}
            \pgfnodelabel{node111121}{node1111211}[0.5][0pt]{\pgfbox[left,center]{\small{3}}}
        \pgfnodecircle{node111122}[stroke]{\pgfxy(2.5,2.75)}{3pt}
          \pgfnodeconnline{node11112}{node111122}
          \pgfnodelabel{node11112}{node111122}[0.5][0pt]{\pgfbox[center,center]{\small{$\#1$}}}
          \pgfnodecircle{node1111221}[stroke]{\pgfxy(2.5,2.0)}{3pt}
            \pgfnodeconnline{node111122}{node1111221}
            \pgfnodelabel{node111122}{node1111221}[0.5][0pt]{\pgfbox[left,center]{\small{2}}}
      \pgfnodecircle{node11113}[stroke]{\pgfxy(3.5,3.5)}{3pt}
        \pgfnodeconnline{node1111}{node11113}
        \pgfnodelabel{node1111}{node11113}[0.5][0pt]{\pgfbox[center,center]{\small{$\#2$}}}
        \pgfnodecircle{node111131}[stroke]{\pgfxy(3.0,2.75)}{3pt}
          \pgfnodeconnline{node11113}{node111131}
          \pgfnodelabel{node11113}{node111131}[0.5][0pt]{\pgfbox[center,center]{\small{$@$}}}
          \pgfnodeconncurve{node111131}{node111}{260}{180}{4.0cm}{5cm}
            \pgfnodelabel{node111131}{node111}[-0.5][1.5cm]{\pgfbox[center,center]{\small{$@(map)$}}}
          \pgfnodecircle{node1111311}[stroke]{\pgfxy(3.5,2.0)}{3pt}
            \pgfnodeconnline{node111131}{node1111311}
            \pgfnodelabel{node111131}{node1111311}[0.5][0pt]{\pgfbox[center,center]{\small{$\#1$}}}
            \pgfnodecircle{node11113111}[stroke]{\pgfxy(4.0,1.25)}{3pt}
              \pgfnodeconnline{node1111311}{node11113111}
              \pgfnodelabel{node1111311}{node11113111}[0.5][0pt]{\pgfbox[left,center]{\small{1}}}
        \pgfnodecircle{node111132}[stroke]{\pgfxy(4.0,2.75)}{3pt}
          \pgfnodeconnline{node11113}{node111132}
          \pgfnodelabel{node11113}{node111132}[0.5][0pt]{\pgfbox[center,center]{\small{$\#1$}}}
          \pgfnodecircle{node1111321}[stroke]{\pgfxy(4.5,2.0)}{3pt}
            \pgfnodeconnline{node111132}{node1111321}
            \pgfnodelabel{node111132}{node1111321}[0.5][0pt]{\pgfbox[left,center]{\small{3}}}
  \pgfnodecircle{node112}[stroke]{\pgfxy(3.5,5.0)}{3pt}
    \pgfnodeconnline{node11}{node112}
    \pgfnodelabel{node11}{node112}[0.5][0pt]{\pgfbox[center,center]{\small{$\#1$}}}
    \pgfnodecircle{node1121}[stroke]{\pgfxy(4.0,4.25)}{3pt}
      \pgfnodeconnline{node112}{node1121}
      \pgfnodelabel{node112}{node1121}[0.5][0pt]{\pgfbox[center,center]{\small{$xs$}}}
\pgfnodecircle{node12}[stroke]{\pgfxy(4.0,5.75)}{3pt}
  \pgfnodeconnline{node1}{node12}
  \pgfnodelabel{node1}{node12}[0.5][0pt]{\pgfbox[center,center]{\small{$\#1$}}}
  \pgfnodecircle{node121}[stroke]{\pgfxy(4.5,5.0)}{3pt}
    \pgfnodeconnline{node12}{node121}
    \pgfnodelabel{node12}{node121}[0.5][0pt]{\pgfbox[center,center]{\small{$f$}}}
\end{pgfpicture}}
\caption{LTS for \textit{map} Skeleton.}
\label{fig:lts_for_map_skeleton}
\end{minipage}
\hfill
\begin{minipage}[b]{0.4\textwidth}
\scalebox{0.99}{
\begin{pgfpicture}{0cm}{0cm}{5cm}{8cm}
\pgfsetendarrow{\pgfarrowto}
\pgfnodecircle{node1}[stroke]{\pgfxy(3.5,7.5)}{3pt}
\pgfnodecircle{node11}[stroke]{\pgfxy(3.0,6.75)}{3pt}
  \pgfnodeconnline{node1}{node11}
  \pgfnodelabel{node1}{node11}[0.5][0pt]{\pgfbox[center,center]{\small{$@$}}}
  \pgfnodecircle{node111}[stroke]{\pgfxy(2.5,6.0)}{3pt}
    \pgfnodeconnline{node11}{node111}
    \pgfnodelabel{node11}{node111}[0.5][0pt]{\pgfbox[center,center]{\small{$@$}}}
    \pgfnodecircle{node1111}[stroke]{\pgfxy(0.0,4.0)}{3pt}
      \pgfnodeconnline{node111}{node1111}
      \pgfnodelabel{node111}{node1111}[0.7][0pt]{\pgfbox[right,center]{\small{$(c_1:w)\ yss$}}}
      \pgfnodecircle{node11111}[stroke]{\pgfxy(0.0,3.25)}{3pt}
        \pgfnodeconnline{node1111}{node11111}
        \pgfnodelabel{node1111}{node11111}[0.5][0pt]{\pgfbox[center,center]{\scriptsize{$[\;]$}}}
    \pgfnodecircle{node1112}[stroke]{\pgfxy(2.0,4.0)}{3pt}
      \pgfnodeconnline{node111}{node1112}
      \pgfnodelabel{node111}{node1112}[0.7][0pt]{\pgfbox[center,center]{\small{$(c_2:w)\ yss$}}}
      \pgfnodecircle{node11121}[stroke]{\pgfxy(2.0,3.25)}{3pt}
        \pgfnodeconnline{node1112}{node11121}
        \pgfnodelabel{node1112}{node11121}[0.5][0pt]{\pgfbox[center,center]{\scriptsize{$[\;]$}}}
    \pgfnodecircle{node1113}[stroke]{\pgfxy(4.5,4.0)}{3pt}
      \pgfnodeconnline{node111}{node1113}
      \pgfnodelabel{node111}{node1113}[0.7][0pt]{\pgfbox[center,center]{\hspace{15mm}\small{$((c_3\ xs\ zs):w)\ yss$}}}
      \pgfnodecircle{node11131}[stroke]{\pgfxy(3.5,3.25)}{3pt}
        \pgfnodeconnline{node1113}{node11131}
        \pgfnodelabel{node1113}{node11131}[0.5][0pt]{\pgfbox[center,center]{\small{$\mathbf{let}$}}}
      \pgfnodecircle{node111311}[stroke]{\pgfxy(2.5,2.5)}{3pt}
        \pgfnodeconnline{node11131}{node111311}
        \pgfnodelabel{node11131}{node111311}[0.5][0pt]{\pgfbox[right,center]{\small{$(:)$}}}
      \pgfnodecircle{node111312}[stroke]{\pgfxy(3.5,2.5)}{3pt}
        \pgfnodeconnline{node11131}{node111312}
        \pgfnodelabel{node11131}{node111312}[0.5][0pt]{\pgfbox[center,center]{\small{$\#1$}}}
        \pgftext[at={\pgfpoint{3.5cm}{2.1cm}}]{\Large{$\ldots$}}
      \pgfnodecircle{node111313}[stroke]{\pgfxy(5.0,2.5)}{3pt}
        \pgfnodeconnline{node11131}{node111313}
        \pgfnodelabel{node11131}{node111313}[0.5][0pt]{\pgfbox[center,center]{\small{$\#2$}}}
        \pgfnodecircle{node1113131}[stroke]{\pgfxy(4.5,1.75)}{3pt}
          \pgfnodeconnline{node111313}{node1113131}
          \pgfnodelabel{node111313}{node1113131}[0.5][0pt]{\pgfbox[center,center]{\small{$@$}}}
          \pgfnodecircle{node11131311}[stroke]{\pgfxy(4.5,1.0)}{3pt}
            \pgfnodeconnline{node1113131}{node11131311}
            \pgfnodelabel{node1113131}{node11131311}[0.5][0pt]{\pgfbox[center,center]{\small{$\#1$}}}
              \pgfnodeconncurve{node11131311}{node111}{210}{180}{6cm}{7cm}
              \pgfnodelabel{node11131311}{node111}[0.1][3.0cm]{\pgfbox[center,center]{\small{$@(mMul'_1)$}}}
            \pgfnodecircle{node111313111}[stroke]{\pgfxy(4.5,0.25)}{3pt}
              \pgfnodeconnline{node11131311}{node111313111}
              \pgfnodelabel{node11131311}{node111313111}[0.5][0pt]{\pgfbox[left,center]{\small{2}}}
        \pgfnodecircle{node1113132}[stroke]{\pgfxy(5.5,1.75)}{3pt}
          \pgfnodeconnline{node111313}{node1113132}
          \pgfnodelabel{node111313}{node1113132}[0.5][0pt]{\pgfbox[center,center]{\small{$\#1$}}}
          \pgfnodecircle{node11131321}[stroke]{\pgfxy(5.5,1.0)}{3pt}
            \pgfnodeconnline{node1113132}{node11131321}
            \pgfnodelabel{node1113132}{node11131321}[0.5][0pt]{\pgfbox[left,center]{\small{1}}}
      \pgfnodecircle{node11132}[stroke]{\pgfxy(5.5,3.25)}{3pt}
        \pgfnodeconnline{node1113}{node11132}
        \pgfnodelabel{node1113}{node11132}[0.5][0pt]{\pgfbox[left,center]{\small{$v$}}}
        \pgftext[at={\pgfpoint{5.5cm}{2.9cm}}]{\Large{$\ldots$}}
  \pgfnodecircle{node112}[stroke]{\pgfxy(3.5,6.0)}{3pt}
    \pgfnodeconnline{node11}{node112}
    \pgfnodelabel{node11}{node112}[0.5][0pt]{\pgfbox[center,center]{\small{$\#1$}}}
    \pgfnodecircle{node1121}[stroke]{\pgfxy(4.0,5.25)}{3pt}
      \pgfnodeconnline{node112}{node1121}
      \pgfnodelabel{node112}{node1121}[0.5][0pt]{\pgfbox[center,center]{\small{$w$}}}
\pgfnodecircle{node12}[stroke]{\pgfxy(4.0,6.75)}{3pt}
  \pgfnodeconnline{node1}{node12}
  \pgfnodelabel{node1}{node12}[0.5][0pt]{\pgfbox[center,center]{\small{$\#1$}}}
  \pgfnodecircle{node121}[stroke]{\pgfxy(4.5,6.0)}{3pt}
    \pgfnodeconnline{node12}{node121}
    \pgfnodelabel{node12}{node121}[0.5][0pt]{\pgfbox[center,center]{\small{$yss$}}}
\end{pgfpicture}}
\caption{LTS for $mMul'_1$ Function.}
\label{fig:lts_for_mmul'_1_function}
\end{minipage}
\end{figure}

\subsection{Parallel Implementation of Skeletons}
\label{sec:parallel_implementation_of_skeletons}
In order to evaluate the parallel programs obtained by our method presented in this chapter, we require efficient parallel implementations of the map and map-reduce skeletons. For the work presented in this paper, we use the Eden library \cite{L2012Eden} that provides parallel implementations of the map and map-reduce skeletons in the following forms:
\begin{center}
{\fontsize{10}{11}\selectfont
\begin{tabular}{@{\hspace{0mm}}l@{\hspace{1mm}}c@{\hspace{1mm}}l@{\hspace{0mm}}}
  \textit{farmB}      & $::$ & $($\textit{Trans} $a,$ \textit{Trans} $b)\ \Rightarrow Int \to (a \to b) \to [a] \to [b]$ \\
  \textit{parMapRedr} & $::$ & $($\textit{Trans} $a,$ \textit{Trans} $b)\ \Rightarrow (b \to b \to b) \to b \to (a \to b) \to [a] \to b$ \\
  \textit{parMapRedl} & $::$ & $($\textit{Trans} $a,$ \textit{Trans} $b)\ \Rightarrow (b \to b \to b) \to b \to (a \to b) \to [a] \to b$
\end{tabular}}
\end{center}

The \textit{farmB} skeleton implemented in Eden divides a given list into \textit{N} sub-lists and creates \textit{N} parallel processes, each of which applies the map computation on a sub-list. The parallel map-reduce skeletons, \textit{parMapRedr} and \textit{parMapRedr}, are implemented using the \textit{parMap} skeleton which applies the \textit{map} operation in parallel on each element in a given list. The result of \textit{parMap} is reduced sequentially using the conventional \textit{foldr} and \textit{foldl} functions, respectively.

Currently, the map-reduce skeletons in the Eden library are defined using the \textit{foldr} and \textit{foldl} functions that require a unit value for the reduction/fold operator to be provided as an input. However, it is evident from Property \ref{pro:non_empty_encoded_list} that the skeletons that are potentially identified will always be applied on non-empty lists. Therefore, we augment the skeletons provided in Eden by adding the following parallel map-reduce skeletons that are defined using the \textit{foldr1} and \textit{foldl1} functions, which are defined for non-empty lists, thereby avoiding the need to obtain a unit value for the reduction operator.

\begin{center}
{\fontsize{10}{11}\selectfont
\begin{tabular}{@{\hspace{0mm}}l@{\hspace{2mm}}c@{\hspace{2mm}}l@{\hspace{0mm}}}
  \textit{parMapRedr1} & $::$ & $($\textit{Trans} $a,$ \textit{Trans} $b)\ \Rightarrow (b \to b \to b) \to (a \to b) \to [a] \to b$ \\
  \textit{parMapRedl1} & $::$ & $($\textit{Trans} $a,$ \textit{Trans} $b)\ \Rightarrow (b \to b \to b) \to (a \to b) \to [a] \to b$
\end{tabular}}
\end{center}

To execute the encoded program produced by our transformation in parallel, we replace the identified skeleton instances with suitable calls to the corresponding skeletons in the Eden library. For example, by replacing functions $mMul'_1$ and $mMul'_3$, which are instances of \textit{map} and \textit{mapReduce} skeletons respectively, with suitable calls to \textit{parMap} and \textit{parMapRedr1}, we obtain the transformed matrix multiplication program $mMul''$ shown in Example \ref{ex:matrix_multiplication_encoded_parallel_program}. \\[3mm]

\begin{example}[Matrix Multiplication -- Encoded Parallel Program]
\label{ex:matrix_multiplication_encoded_parallel_program}
{\fontsize{10}{11}\selectfont
\begin{tabular}{@{\hspace{0mm}}l@{\hspace{2mm}}c@{\hspace{2mm}}l@{\hspace{0mm}}}
  \multicolumn{3}{@{\hspace{0mm}}l@{\hspace{0mm}}}{$mMul''\ xss\ yss$} \\
  \multicolumn{3}{@{\hspace{0mm}}l@{\hspace{0mm}}}{\textbf{where}} \\
  $mMul''\ xss\ yss$ & $=$ & $mMul''_1\ (encode_{mMul_1}\ xss\ yss)\ yss$ \\[0mm]
  $mMul''_1\ w\ yss$ & $=$ & \textit{farmB} $\ noPe\ f\ w$ \\
                     &     & \textbf{where} \\
                     &     & \begin{tabular}{@{\hspace{0mm}}l@{\hspace{2mm}}c@{\hspace{2mm}}l@{\hspace{0mm}}}
                               $f\ c_1$ & $=$ & $[]$ \\
                               $f\ c_2$ & $=$ & $[]$ \\
                               $f\ (c_3\ xs\ zs)$ & $=$ & $\mathbf{let}\ v = \lambda xs. g\ xs$ \\
                                                  &     & \begin{tabular}{@{\hspace{17mm}}l@{\hspace{2mm}}c@{\hspace{2mm}}l@{\hspace{0mm}}}
                                                            \multicolumn{3}{@{\hspace{17mm}}l@{\hspace{0mm}}}{$\mathbf{where}$} \\
                                                            $g\ []$ & $=$ & $0$ \\
                                                            $g\ (x:xs)$ & $=$ & $x$
                                                          \end{tabular} \\
                                                  &     & $\mathbf{in}\ mMul''_2\ (encode_{mMul_2}\ zs)\ xs\ yss\ v$
                             \end{tabular}
\end{tabular}

\noindent
\begin{tabular}{@{\hspace{0mm}}l@{\hspace{2mm}}c@{\hspace{2mm}}l@{\hspace{0mm}}}
  $mMul''_2\ (c_4:w)\ xs\ yss\ v$ & $=$ & $[]$ \\
  $mMul''_2\ (c_5:w)\ xs\ yss\ v$ & $=$ & $\mathbf{let}\ v' = \lambda xs. g\ xs$ \\
                                  &     & \begin{tabular}{@{\hspace{18mm}}l@{\hspace{2mm}}c@{\hspace{2mm}}l@{\hspace{0mm}}}
                                            \multicolumn{3}{@{\hspace{18mm}}l@{\hspace{0mm}}}{$\mathbf{where}$} \\
                                            $g\ []$ & $=$ & $0$ \\
                                            $g\ (x:xs)$ & $=$ & $v\ xs$
                                          \end{tabular} \\
                                  &     & $\mathbf{in}\ (mMul''_3\ (encode_{mMul_3}\ xs\ yss)\ v):(mMul''_2\ w\ xs\ yss\ v')$
\end{tabular}

\noindent
\begin{tabular}{@{\hspace{0mm}}l@{\hspace{2mm}}c@{\hspace{2mm}}l@{\hspace{0mm}}}
  $mMul''_3\ w\ v$ & $=$ & \textit{parMapRedr1} $g\ f\ w$ \\
                   &     & \textbf{where} \\
                   &     & \begin{tabular}{@{\hspace{0mm}}l@{\hspace{2mm}}c@{\hspace{2mm}}l@{\hspace{0mm}}}
                             $g\ x\ y$ & $=$ & $x + y$ \\
                             $f\ c_6$ & $=$ & $0$ \\
                             $f\ c_7$ & $=$ & $0$ \\
                             $f\ (c_8\ x\ ys)$ & $=$ & $x * (v\ ys)$
                           \end{tabular}
\end{tabular}}
\end{example}

\vspace{2mm}
\noindent
\textbf{Consequence:} By automatically identifying instances of list-based map and map-reduce skeletons, we produce a program that is defined using these parallelisable skeletons. Using parallel implementations for these skeletons that are available in existing libraries such as Eden, it is possible to execute the transformed program on parallel hardware.

\section{Evaluation}
\label{sec:evaluation}
In this paper, we present the evaluation of two benchmark programs -- matrix multiplication and dot-product of binary trees -- to illustrate interesting aspects of our transformation. The programs are evaluated on a Mac Pro computer with a 12-core Intel Xeon E5 processor each clocked at 2.7 GHz and 64 GB of main memory clocked at 1866 MHz. GHC version 7.10.2 is used for the sequential versions of the benchmark programs and the latest Eden compiler based on GHC version 7.8.2 for the parallel versions.

For all parallel versions of a benchmark program, only those skeletons that are present in the top-level expression are executed using their parallel implementations. That is, nesting of parallel skeletons is avoided. The nested skeletons that are present inside top-level skeletons are executed using their sequential versions. The objective of this approach is to avoid uncontrolled creation of too many threads which we observe to result in inefficient parallel execution where the cost of thread creation and management is greater than the cost of parallel execution.

\subsection{Example -- Matrix Multiplication}
\label{sec:example_matrix_multiplication}
The original sequential version, distilled version, encoded version and encoded parallel version of the matrix multiplication program are presented in Examples \ref{ex:matrix_multiplication_original_program}, \ref{ex:matrix_multiplication_distilled_program}, \ref{ex:matrix_multiplication_encoded_program} and \ref{ex:matrix_multiplication_encoded_parallel_program} respectively.

A hand-parallel version of the original matrix multiplication program in Example \ref{ex:matrix_multiplication_original_program} is presented in Example \ref{ex:matrix_multiplication_hand_parallel_program}. We identify that function \textit{mMul} is an instance of the \textit{map} skeleton and therefore define it using a suitable call to the \textit{farmB} skeleton available in the Eden library.

\begin{example}[Matrix Multiplication -- Hand-Parallel Program]
\label{ex:matrix_multiplication_hand_parallel_program}
{\fontsize{10}{11}\selectfont
\begin{tabular}{@{\hspace{0mm}}l@{\hspace{2mm}}c@{\hspace{2mm}}l@{\hspace{0mm}}}
  \multicolumn{3}{@{\hspace{0mm}}l@{\hspace{0mm}}}{$mMul\ xss\ yss$} \\
  \multicolumn{3}{@{\hspace{0mm}}l@{\hspace{0mm}}}{$\mathbf{where}$} \\
  $mMul\ xss\ yss$ & $=$ & $farmB\ noPe\ f\ xss$ \\
                   &     & \textbf{where} \\
                   &     & $f\ xs\ =\ map\ (dotp\ xs)\ (transpose\ yss)$ \\[1mm]
  $dotp\ xs\ ys$ & $=$ & $foldr\ (+)\ 0\ (zipWith\ (*)\ xs\ ys)$ \\[1mm]
  $transpose\ xss$ & $=$ & $transpose'\ xss\ []$ \\
  $transpose'\ []\ yss$ & $=$ & $yss$ \\
  $transpose'\ (xs:xss)\ yss$ & $=$ & $transpose'\ xss\ (rotate\ xs\ yss)$ \\[1mm]
  $rotate\ []\ yss$ & $=$ & $yss$ \\
  $rotate\ (x:xs)\ []$ & $=$ & $[x]:(rotate\ xs\ yss)$ \\
  $rotate\ (x:xs)\ (ys:yss)$ & $=$ & $(ys \doubleplus [x]):(rotate\ xs\ yss)$
\end{tabular}}
\end{example}

\begin{figure}[!ht]
\centering
\includegraphics[scale=0.40]{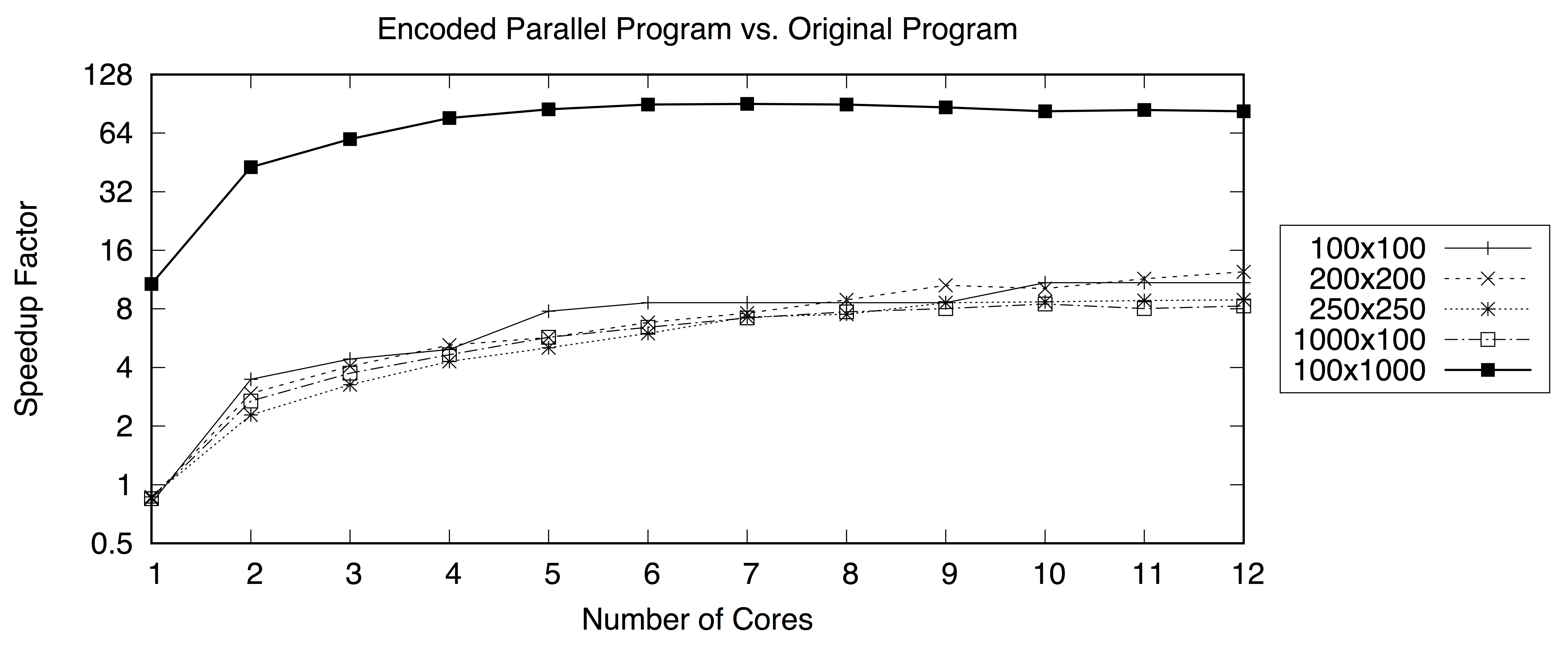}
\includegraphics[scale=0.40]{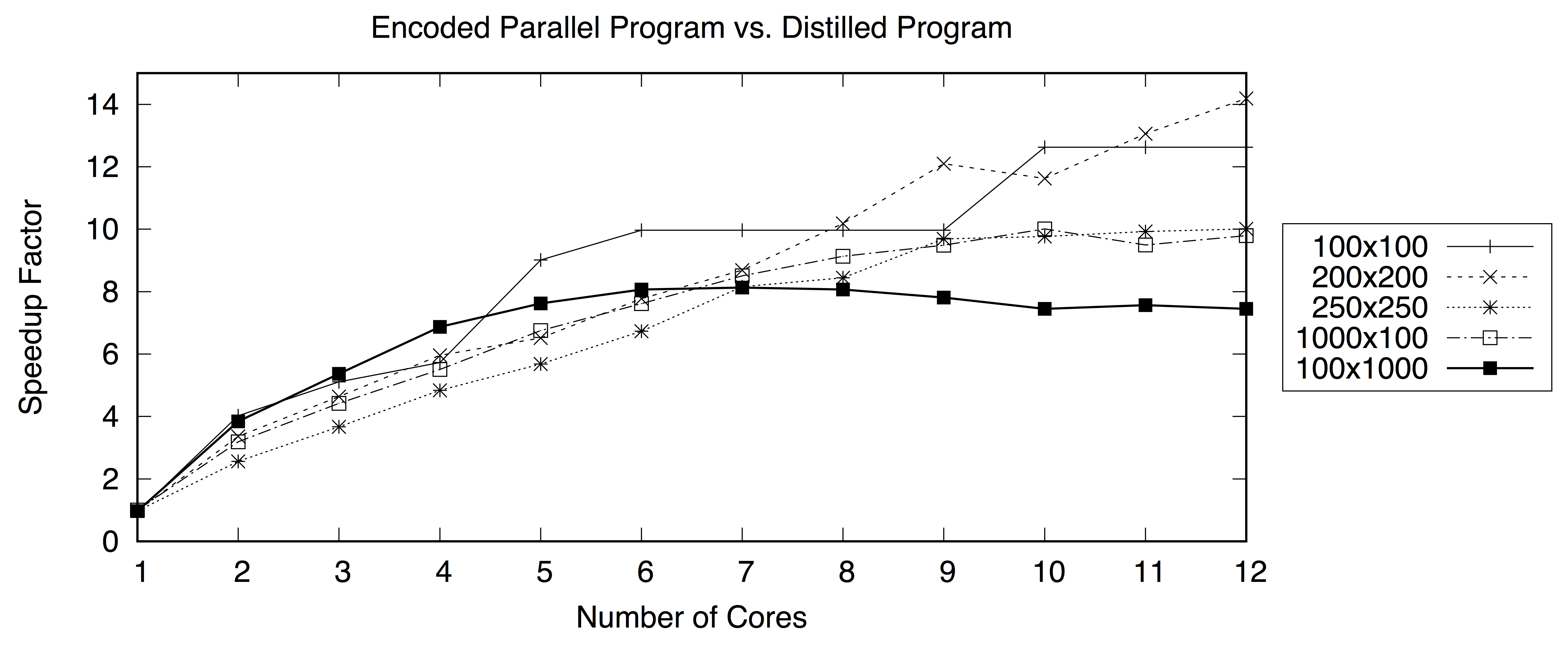}
\includegraphics[scale=0.40]{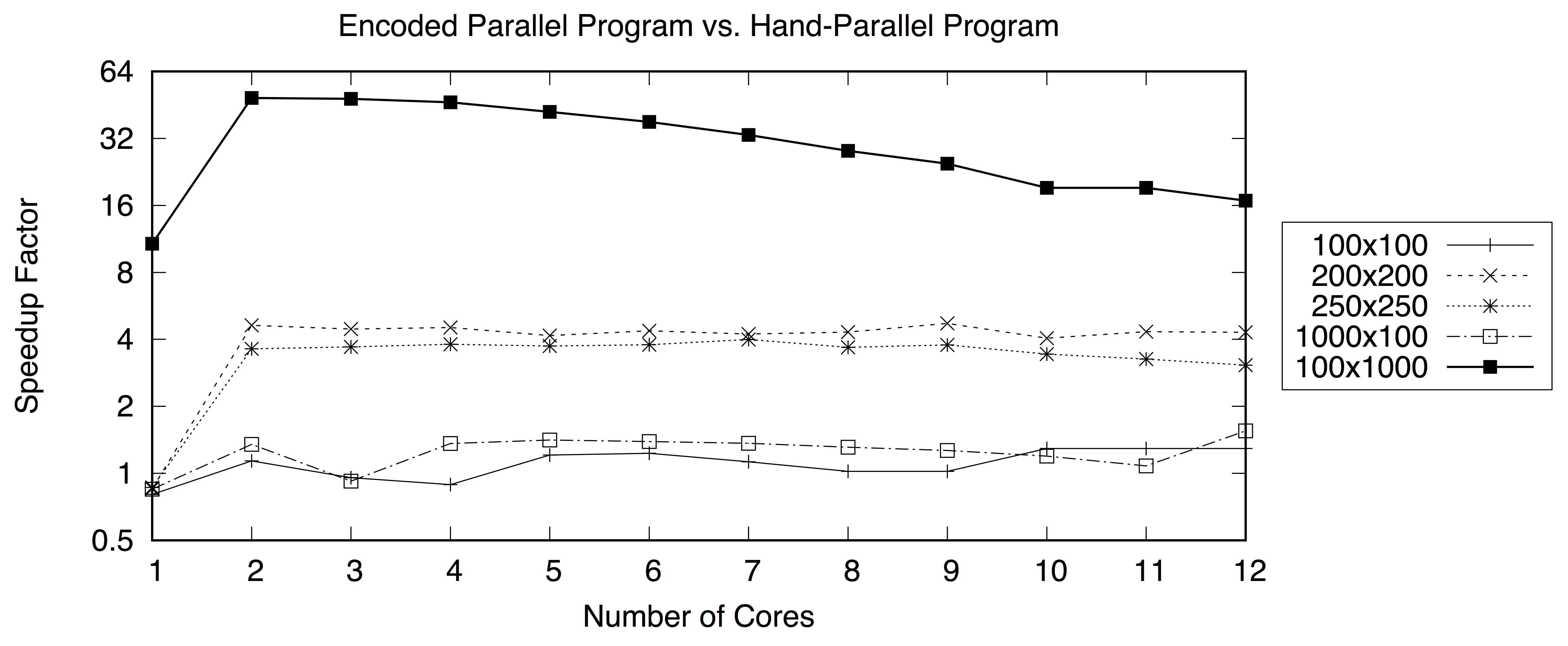}
\caption{Evaluation of Speedup for Matrix Multiplication}
\label{fig:evaluation_of_speedup_for_matrix_multiplication}
\end{figure}

Figure \ref{fig:evaluation_of_speedup_for_matrix_multiplication} presents the speedups achieved by the encoded parallel version of the matrix multiplication program in comparison with the original, distilled and hand-parallel versions. Since we avoid nested parallel skeletons as explained earlier, the encoded parallel program used in this evaluation contains only the $mMul''_1$ function defined using the \textit{farmB} skeleton and uses the sequential definition for $mMul''_3$. An input size indicated by \texttt{NxM} denotes the multiplication of matrices of sizes \texttt{NxM} and \texttt{MxN}.

When compared to the original program, we observe that the encoded parallel version achieves a positive speedup of \texttt{3x-8x} for all input sizes except for \texttt{100x1000}. In the case with input size \texttt{100x1000}, the speedup achieved is \texttt{6x-20x} more than the speedups achieved for the other input sizes. This is due to the intermediate data structure \textit{transpose yss}, which is of the order of 1000 elements for input size \texttt{100x1000} and of the order of 100 elements for the other inputs, that is absent in the encoded parallel program. This can be verified from the comparison with the distilled version, which is also free of intermediate data structures. Hence, the encoded parallel program has a linear speedup compared to the distilled version.
From Examples \ref{ex:matrix_multiplication_hand_parallel_program} and \ref{ex:matrix_multiplication_encoded_parallel_program}, we observe that both the hand-parallel and encoded parallel versions parallelise the equivalent computations that multiply rows in the first matrix with columns in the second matrix using the \textit{farmB} skeleton. However, the encoded parallel version is marginally faster than the hand-parallel version for input sizes \texttt{100x100} and \texttt{1000x100}, and \texttt{4x} faster for other input sizes except \texttt{100x1000}. This is due to the use of intermediate data structures in the hand-parallel version which is of the order of \texttt{100} for all input sizes except \texttt{100x1000} for which the speedup achieved is \texttt{48x-18x} more than the hand-parallel version. Also, the hand-parallel version scales better with a higher number of cores than the encoded parallel version for the input size \texttt{100x1000}. This is because the encoded parallel version achieves better speedup even with fewer cores due to the elimination of intermediate data structures, and hence does not scale as impressively as the hand-parallel version.

\subsection{Example -- Dot-Product of Binary Trees}
\label{sec:example_dot_product_of_binary_trees}
Example \ref{ex:dot_product_of_binary_trees_original_distilled_program} presents a sequential program to compute the dot-product of binary trees, where \textit{dotP} computes the product of values at the corresponding branch nodes of trees $xt$ and $yt$, and adds the dot-products of the left and right sub-trees. The distilled version of this program remains the same as there are no intermediate data structures and a hand-parallel version cannot be defined using list-based parallel skeletons.
\begin{example}[Dot-Product of Binary Trees -- Original/Distilled Program]
\label{ex:dot_product_of_binary_trees_original_distilled_program}
{\fontsize{10}{11}\selectfont
\begin{tabular}{@{\hspace{0mm}}l@{\hspace{1mm}}c@{\hspace{1mm}}l@{\hspace{0mm}}}
  \multicolumn{3}{@{\hspace{0mm}}l@{\hspace{0mm}}}{$\mathbf{data}\ BTree\ a\ ::=\ E\ |\ B\ a\ (BTree\ a)\ (BTree\ a)$} \\[1mm]
  \multicolumn{3}{@{\hspace{0mm}}l@{\hspace{0mm}}}{$dotP\ ::\ (BTree\ a) \to (BTree\ a) \to (BTree\ a)$} \\
  \multicolumn{3}{@{\hspace{0mm}}l@{\hspace{0mm}}}{$dotP\ xt\ yt$} \\
  \multicolumn{3}{@{\hspace{0mm}}l@{\hspace{0mm}}}{\textbf{where}} \\
  $dotP\ E\ yt$ & $=$ & $0$ \\
  $dotP\ (B\ x\ xt_1\ xt_2)\ E$ & $=$ & $0$ \\
  $dotP\ (B\ x\ xt_1\ xt_2)\ (B\ y\ yt_1\ yt_2)$ & $=$ & $(x*y) + (dotP\ xt_1\ yt_1) + (dotP\ xt_2\ yt_2)$
\end{tabular}}
\end{example}

By applying the encoding transformation, we obtain the encoded version for the dot-product program as shown in Example \ref{ex:dot_product_of_binary_trees_original_encoded_program}.
\begin{example}[Dot-Product of Binary Trees -- Encoded Program]
\label{ex:dot_product_of_binary_trees_original_encoded_program}
{\fontsize{10}{11}\selectfont
\begin{tabular}{@{\hspace{0mm}}l@{\hspace{2mm}}c@{\hspace{2mm}}l@{\hspace{0mm}}}
  $\mathbf{data}\ T_{dotP}\ a$ & $::=$ & $c_1\ |\ c_2\ |\ c_3\ a\ a\ (BTree\ a)\ (BTree\ a)$
\end{tabular}

\vspace{1mm}
\noindent
\begin{tabular}{@{\hspace{0mm}}l@{\hspace{2mm}}c@{\hspace{2mm}}l@{\hspace{0mm}}}
  $encode_{dotP}\ E\ yt$ & $=$ & $[c_1]$ \\
  $encode_{dotP}\ (B\ x\ xt_1\ xt_2)\ E$ & $=$ & $[c_2]$ \\
  $encode_{dotP}\ (B\ x\ xt_1\ xt_2)\ (B\ y\ yt_1\ yt_2)$ & $=$ & $[c_3\ x\ y\ xt_1\ yt_1] \doubleplus (encode_{dotP}\ xt_2\ yt_2)$ \\
\end{tabular}

\vspace{1mm}
\noindent
\begin{tabular}{@{\hspace{0mm}}l@{\hspace{2mm}}c@{\hspace{2mm}}l@{\hspace{0mm}}}
  \multicolumn{3}{@{\hspace{0mm}}l@{\hspace{0mm}}}{$dotP\ (encode_{dotP}\ xt\ yt)$} \\
  \multicolumn{3}{@{\hspace{0mm}}l@{\hspace{0mm}}}{\textbf{where}} \\
  $dotP\ (c_1 : w)$ & $=$ & $0$ \\
  $dotP\ (c_2 : w)$ & $=$ & $0$ \\
  $dotP\ \big((c_3\ x\ y\ xt\ yt) : w\big)$ & $=$ & $(x*y) + (dotP\ (encode_{dotP}\ xt\ yt)) + (dotP\ w)$ \\[2mm]
\end{tabular}}
\end{example}

By applying the skeleton identification rule to this encoded program, we identify that the encoded version of function \textit{dotP} is an instance of the \textit{mapRedr} skeleton. Example \ref{ex:dot_product_of_binary_trees_encoded_parallel_program} shows the encoded parallel version defined using a suitable call to the \textit{parMapRedr1} skeleton in the Eden library. As explained before, we define only the top-level call to $dotP''$ using the parallel skeleton and use the sequential for the nested call to \textit{dotP} because we avoid nested parallel skeletons in this evaluation.

\begin{example}[Dot Product of Binary Trees -- Encoded Parallel Program]
\label{ex:dot_product_of_binary_trees_encoded_parallel_program}
{\fontsize{10}{11}\selectfont
\begin{tabular}{@{\hspace{0mm}}l@{\hspace{1mm}}c@{\hspace{1mm}}l@{\hspace{0mm}}}
  $\mathbf{data}\ T_{dotP}\ a$ & $::=$ & $c_1\ |\ c_2\ |\ c_3\ a\ a\ (BTree\ a)\ (BTree\ a)$ \\[1mm]
\end{tabular}}

\vspace{2mm}
\noindent
{\fontsize{10}{11}\selectfont
\begin{tabular}{@{\hspace{0mm}}l@{\hspace{1mm}}c@{\hspace{1mm}}l@{\hspace{0mm}}}
  $encode_{dotP}\ E\ yt$ & $=$ & $[c_1]$ \\
  $encode_{dotP}\ (B\ x\ xt_1\ xt_2)\ E$ & $=$ & $[c_2]$ \\
  $encode_{dotP}\ (B\ x\ xt_1\ xt_2)\ (B\ y\ yt_1\ yt_2)$ & $=$ & $[c_3\ x\ y\ xt_1\ yt_1] \doubleplus (encode_{dotP}\ xt_2\ yt_2)$
\end{tabular}}

\vspace{2mm}
\noindent
{\fontsize{10}{11}\selectfont
\begin{tabular}{@{\hspace{0mm}}l@{\hspace{1mm}}c@{\hspace{1mm}}l@{\hspace{0mm}}}
  \multicolumn{3}{@{\hspace{0mm}}l@{\hspace{0mm}}}{$dotP''\ (encode_{dotP}\ xt\ yt)$} \\
  \multicolumn{3}{@{\hspace{0mm}}l@{\hspace{0mm}}}{\textbf{where}} \\
  $dotP''\ w$ & $=$ & \textit{parMapRedr1} $g\ f\ w$ \\
                         &     & \textbf{where} \\
                         &     & \begin{tabular}{@{\hspace{0mm}}l@{\hspace{2mm}}c@{\hspace{2mm}}l@{\hspace{0mm}}}
                                   $g$ & $=$ & $(+)$ \\
                                   $f\ c_1$ & $=$ & $0$ \\
                                   $f\ c_2$ & $=$ & $0$ \\
                                   $f\ (c_3\ x\ y\ xt\ yt)$ & $=$ & $(x*y) + (dotP\ xt\ yt)$
                                 \end{tabular}
\end{tabular}}

\vspace{2mm}
\noindent
{\fontsize{10}{11}\selectfont
\begin{tabular}{@{\hspace{0mm}}l@{\hspace{1mm}}c@{\hspace{1mm}}l@{\hspace{0mm}}}
  $dotP\ E\ yt$ & $=$ & $0$ \\
  $dotP\ (B\ x\ xt_1\ xt_2)\ E$ & $=$ & $0$ \\
  $dotP\ (B\ x\ xt_1\ xt_2)\ (B\ y\ yt_1\ yt_2)$ & $=$ & $(x*y) + (dotP\ xt_1\ yt_1) + (dotP\ xt_2\ yt_2)$
\end{tabular}}
\end{example}

\begin{figure}[!ht]
\centering
\includegraphics[scale=0.40]{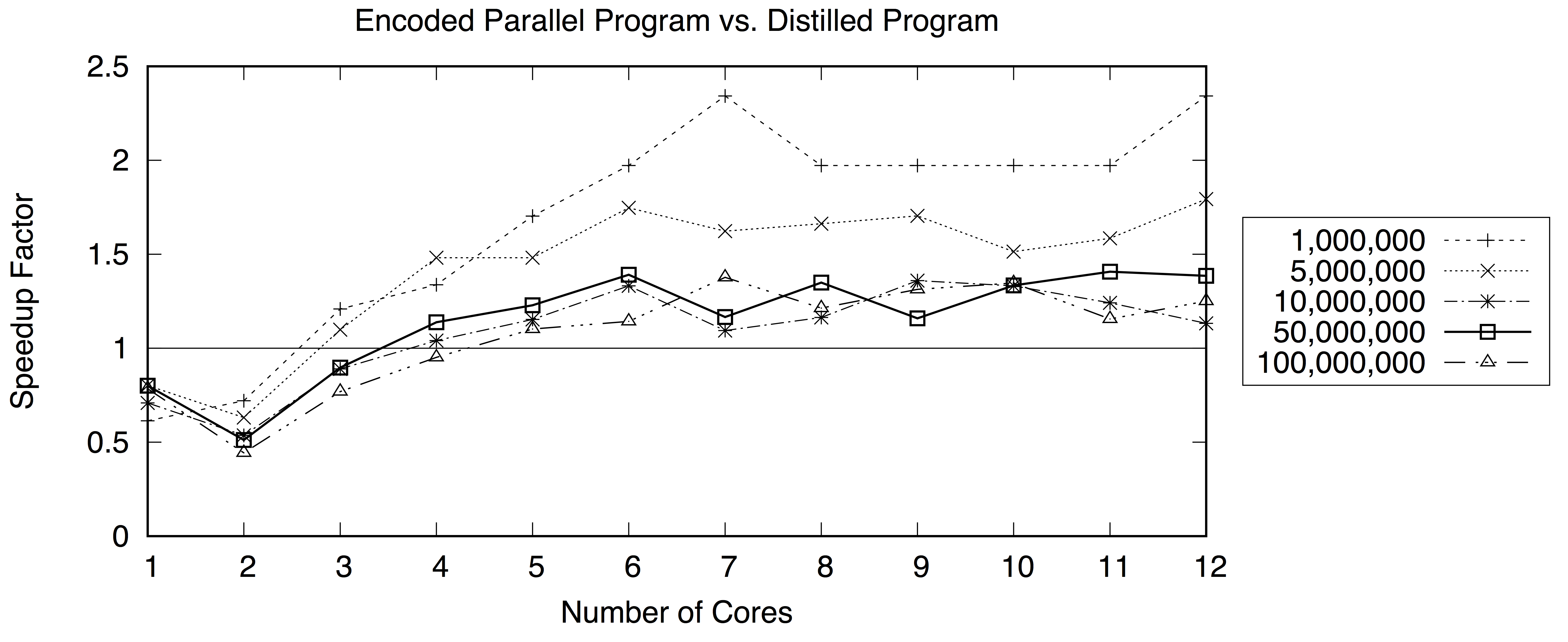}
\caption{Evaluation of Speedup for Dot-Product of Binary Trees}
\label{fig:evaluation_of_speedup_for_dot_product_of_binary_trees}
\end{figure}

Figure \ref{fig:evaluation_of_speedup_for_dot_product_of_binary_trees} presents the speedups of the encoded parallel version compared to the original version. An input size indicated by \texttt{N} denotes the dot-product of two identical balanced binary trees with \texttt{N} nodes each. We observe that the encoded parallel version achieves a positive speedup only upon using more than 4 cores, resulting in a maximum speedup of \texttt{2.4x} for input size \texttt{1,000,000} and \texttt{1.4x} for input size \texttt{100,000,000}. For all input sizes, the speedup factor does not improve when using more than 6 cores. Further, we also observe that the speedup achieved scales negatively as the input size increases.
The reason for this performance of the encoded parallel version is as follows: From Example \ref{ex:dot_product_of_binary_trees_encoded_parallel_program}, we observe that each element in the encoded list contains the values at the branch nodes ($x$ and $y$) and the sub-trees ($xt_1$ and $yt_1$), which are arguments of the first recursive call $dotP\ xt_1\ yt_1$ in Example \ref{ex:dot_product_of_binary_trees_original_distilled_program}. Consequently, the sizes of the elements in the encoded list progressively decrease from the first to the last element if the input trees are balanced. The encoded list is then split into sub-lists in round-robin fashion by the \textit{parMapRedr1} skeleton and each thread in the parallel execution applies the sequential dot-product computation $dotP$ over a sub-list. As a result, the workloads of the parallel threads are not well-balanced and this results in significant idle times for threads that process smaller sub-lists. We also note that, left-skewed input binary trees would result in poorer performance, while right-skewed input binary trees result in better performance of the encoded parallel versions.

\section{Conclusion}
\label{sec:conclusion}

\subsection{Summary}
\label{sec:summary}
We have presented a transformation method to efficiently parallelise a given program by automatically identifying parallel skeletons and reducing the number of intermediate data structures used. By encoding the inputs of a program into a single input, we facilitate the identification of parallel skeletons, particularly for map- and reduce-based computations. Additionally, we can automatically check skeleton operators for desired properties thereby allowing complete automation of the parallelisation process. Importantly, our transformation does not place any restriction on the programs that can be transformed using our method.

To evaluate our transformation method, we presented two interesting benchmark programs whose inputs are encoded into a \textit{cons}-list. From the results, we observe two possible extreme performances. In one case, linear to super-linear speedups are achieved due to the distillation transformation, which reduces the use of intermediate data structures, as well as our parallelisation transformation. In another case, despite parallelising a program that cannot be defined using existing skeleton implementations in libraries, the positive speedups achieved are limited and not as desired. Despite not being discussed here, by employing additional skeletons such as \textit{accumulate} \cite{IH2004ANewParallelSkeleton}, we are able to automatically parallelise interesting programs such as maximum prefix sum.

The primary challenge lies in the efficient execution of the parallel programs produced that are defined using skeletons. It is important to have efficient implementations of these parallel skeletons that incorporate intelligent data-partitioning and load-balancing methods across the parallel threads created to execute the skeletons. We believe that better load-balancing across threads can be facilitated by polytypic parallel skeletons, list or array data structures that support nested parallelism, or dynamic load-balancing at run-time.

\subsection{Related Work}
\label{sec:related_work}
Previously, following the seminal works by Cole \cite{C1992AlgorithmicSkeletons} and Darlington et. al. \cite{DFHKSW1993ParallelProgrammingUsing} on skeleton-based program development, a majority of the work that followed \cite{MIEH2006ALibraryOfConstructiveSkeletons, MKIHA2004AFusionEmbeddedSkeletonLibrary, MHT2006SkeletonsForGeneralTrees, CKLMG2011Accelerate} catered to manual parallel programming. To address the difficulties in choosing appropriate skeletons for a given algorithm, Hu et. al. proposed the \textit{diffusion} transformation \cite{HTI1999DiffusionCalculating}, which is capable of decomposing recursive functions of a certain form into several functions, each of which can be described by a skeleton. Even though diffusion can transform a wider range of functions to the required form, this method is only applicable to functions with one recursive input. Further they proposed the accumulate skeleton \cite{IH2004ANewParallelSkeleton} that encapsulates the computational forms of map and reduce skeletons that use an accumulating parameter to build the result. However, the associative property of the reduce and scan operators used in the accumulate skeleton have to be verified and their unit values derived manually.

The calculational approaches to program parallelisation are based on list-homomorphisms \cite{Skillikorn1993BirdMeertensParallelModel} and propose systematic ways to derive parallel programs. However, most methods are restricted to programs that are defined over lists \cite{G1995ConstructingListHomomorphisms, G1996TheThirdHomomorphism, G1999ExtractingAndImplementingListHomomorphisms, HYT2005ProgramOptimisationsAnd}. Further, they require manual derivation of operators or their verification for certain algebraic properties to enable parallel evaluation of the programs obtained. Morihata et. al. \cite{MMHT2009TheThirdHomomorphism} extended this approach for trees by decomposing a binary tree into a list of sub-trees called \textit{zipper}, and defining upward and downward computations on the zipper structure. However, such calculational methods are often limited by the range of programs and data types they can transform. Also, a common aspect of these calculational approaches is the need to manually derive operators that satisfy certain properties, such as associativity to guarantee parallel evaluation. To address this, Chin et. al. \cite{CH1998ParallelisationViaContextPreservation} proposed a method that systematically derives parallel programs from sequential definitions and automatically creates auxiliary functions that can be used to define associative operators needed for parallel evaluation. However, their method is restricted to a first-order language and applicable to functions defined over a single recursive linear data type, such as lists, that has an associative decomposition operator, such as $\doubleplus$.

As an alternative to calculational approaches, Ahn et. al. \cite{AH2001AnalyticalMethodRecursiveFunctions} proposed an analytical method to transform general recursive functions into a composition of polytypic data parallel skeletons. Even though their method is applicable to a wider range of problems and does not need associative operators, the transformed programs are defined by composing skeletons and employ multiple intermediate data structures.

Previously, the authors proposed a method to transform the input of a given program into a \textit{cons}-list based on the recursive structure of the input \cite{KH2014ExtractingDataParallelComputationsFromDistilledPrograms}. Since this method does not use the recursive structure of the program to build the \textit{cons}-list, the transformed programs do not lend themselves to be defined using list-based parallel skeletons. This observation led to creating a new encoded data type that matches the algorithmic structure of the program and hence enables identification of polytypic parallel map and reduce skeletons \cite{KH2016ProgramTransformationToIdentifyParallelSkeletons}. The new encoded data type is created by pattern-matching and recursively consuming inputs, where a recursive components is created in the new encoded input for each recursive call that occurs in a function body using the input arguments of the recursive call. Consequently, the data structure of the new encoded input reflects the recursive structure of the program. Even though this method leads to better identification of polytypic skeletons, it is not easy to evaluate the performance of the transformed programs defined using these skeletons because existing libraries do not offer implementations of skeletons that are defined over a generic data type. Consequently, the proposed method of encoding the inputs into a list respects the recursive structures of programs and allows evaluation of the transformed programs using existing implementations of list-based parallel skeletons.

\section*{Acknowledgment}
This work was supported, in part, by the Science Foundation Ireland grant 10/CE/I1855 to Lero - the Irish Software Research Centre (www.lero.ie).


\bibliography{mybib}{}
\bibliographystyle{eptcs}

\end{document}